\documentclass[twocolumn,superscriptaddress,floats,prd,nofootinbib,longbibliography]{revtex4-1}
\usepackage{amssymb,amsmath,amssymb,amsfonts,amsthm,stmaryrd,mathrsfs,physics,bbm}
\usepackage{subfigure}
\usepackage{graphicx}
\usepackage[T1]{fontenc}
\usepackage{enumerate} 
\usepackage{color} 
\usepackage{graphpap} 
\usepackage{xcolor}

\usepackage{hyperref} 

\newcommand{\heff}{{H_{\rm eff}}}

\newcommand{\tx}{\mathfrak{x}}

\newcommand{\tp}{\mathfrak{p}}

\newcommand{\BNU}{School of Physics and Astronomy, \mbox{Key Laboratory of Multiscale Spin Physics (Ministry of Education)}, Beijing Normal University, Beijing 100875, China}
\newcommand{\UOE}{Department Physik, Institut f\"ur Quantengravitation, Theoretische Physik III, Friedrich-Alexander-Universit\"at Erlangen-N\"urnberg, Staudtstra{\ss}e 7/B2, 91058 Erlangen, Germany}

\begin{document}

\title{Dust shell in effective loop quantum black hole model}
\author{Hanno Sahlmann}
\email{hanno.sahlmann@gravity.fau.de}
\affiliation{\UOE}

\author{Cong Zhang}
\email{Corresponding author: cong.zhang@bnu.edu.cn}
\affiliation{\BNU}

\begin{abstract}
In this work, the dynamics of a dust shell in an effective theory of spherically symmetric gravity containing quantum corrections from loop quantum gravity is investigated. To provide a consistent framework for including the dust, we go beyond the standard formulation of the effective theory by introducing an action that includes not only the effective Hamiltonian constraint, but also the diffeomorphism constraint, along with appropriate gauge-fixing and boundary terms. By adding the dust shell action and substituting vacuum solutions for the interior and exterior regions, we derive a reduced action in which only the shell radius $\tx$ and the exterior black hole mass appear as dynamical variables. Varying the reduced action yields the evolution equation for $\tx$, which is then solved numerically to explore the dynamical properties of the dust shell and the continuity properties of the metric. Finally, our approach is compared with previous studies to highlight key differences and improvements.
\end{abstract}


\maketitle

\section{introduction}\label{sec:intro}
It is widely expected that quantum gravity (QG) should be incorporated to properly understand the formation and ultimate fate of black holes (BHs). 
Loop quantum gravity \cite{Ashtekar:2004eh,Rovelli:2004tv,Han:2005km,Thiemann:2007pyv}, as a background-independent and non-perturbative approach to quantum gravity, has been applied to the study of BHs \cite{Modesto:2008im,Gambini:2008dy,Corichi:2015xia,BenAchour:2016brs,Ashtekar:2018lag,Zhang:2020qxw,Kelly:2020uwj,Zhang:2021wex,Husain:2021ojz,Alonso-Bardaji:2021yls,Zhang:2021xoa,Munch:2021oqn,Gambini:2022dec,Husain:2022gwp,Alonso-Bardaji:2022ear,Gambini:2022hxr,Lewandowski:2022zce,Alonso-Bardaji:2023vtl,Giesel:2023hys,Giesel:2023tsj,AlonsoBardaji:2023bww,Fazzini:2023ova,Zhang:2024khj,Cafaro:2024vrw,Lin:2024flv,Cipriani:2024nhx,Zhang:2024ney,Lin:2024beb,Yang:2025ufs,Fazzini:2025ysd,Liu:2025fil,Fazzini:2025zrq}. Beyond extensive work on vacuum solutions in loop quantum BH models, recent attention has turned to their formation throughough dust collapse \cite{Husain:2021ojz,Husain:2022gwp,Lewandowski:2022zce,Fazzini:2023ova,Giesel:2023hys,Giesel:2023tsj,Cipriani:2024nhx,Shi:2024vki,Fazzini:2025ysd,Liu:2025fil,Fazzini:2025zrq}. A central issue in these studies is the appearance of shell-crossing singularities. To understand this, let us view the collapsing dust ball as a collection of concentric shells. Due to quantum gravity effects, each shell may undergo a bounce at a different time, depending on the initial density profile. This temporal mismatch leads to collisions between neighboring shells. As these collisions accumulate, they ultimately result in the formation of a thin shell with divergent energy density, referred to as a dust shell (see, e.g., \cite{Fazzini:2023ova,Cipriani:2024nhx,Fazzini:2025ysd,Liu:2025fil,Fazzini:2025zrq} for more details).  Although the existence of the shell-crossing singularity has been widely accepted, the dynamics of the dust shell is still an open issue.  

In works such as \cite{Husain:2021ojz,Cipriani:2024nhx,Liu:2025fil}, the dynamics of the dust shell is studied by finding weak solutions to the equations of motion. This is a standard procedure as done in classical GR. However, a priori assumption in those works differ from that in classical GR. In classical GR, it is assumed that the metric is continuous while its derivatives may be discontinuous \cite{Israel:1966rt,Friedman:1997fu,Casado-Turrion:2023jba}. This ensures the continuity of the induced metric on the world sheet of the dust shell and allows for a discontinuity in the extrinsic curvature. 
However, in those works,  though not explicitly stated, the assumption is that a specific type of coordinates system, Painlev\'e–Gullstrand (PG) coordinates, remain valid coordinates across the shell\footnote{In the works, e.g., \cite{Husain:2021ojz,Cipriani:2024nhx}, when the Rankine-Hugoniot condition is derived, the PG coordinates in the exterior and interior regions are used and assumed to be continuous (see also \cite{Fazzini:2025zrq} for a discussion of this issue).}. In fact, within classical GR, it can be shown that the $t$-coordinates in both the PG chart and the Schwarzschild chart are generally discontinuous at the shell \cite{Israel:1966rt,Friedman:1997fu}. 

Imposing an assumption on a specific coordinate system in a theory that describes spacetime geometry appears unnatural. However, in some loop quantum BH models, this assumption arises as a consequence of limitations inherent in the models themselves. Specifically, before quantizing the classical theory, a gauge-fixing condition is imposed to solve the diffeomorphism constraint \cite{Kelly:2020uwj,Husain:2022gwp}.  As a result, the quantum theory loses spacetime covariance \cite{Kelly:2020uwj,Zhang:2024khj}. Consequently, a specific gauge must be selected to define the spacetime metric \cite{Han:2022rsx}. In work such as \cite{Husain:2021ojz,Cipriani:2024nhx}, the chosen gauge, called the PG gauge, corresponds to the PG coordinates, as this choice naturally leads to dynamics with respect to dust proper time when dust coupling is included. For discussions on restoring general covariance in (loop) quantum BH models, see, for example, \cite{AlonsoBardaji:2023bww,Alonso-Bardaji:2023vtl,Zhang:2024khj,Belfaqih:2024vfk,Zhang:2024ney}.

In these loop quantum black hole models, a matter field, typically a dust field, is introduced for gauge fixing. One begins with the full action of gravity coupled to dust, and then imposes the areal gauge and dust time gauge to solve the diffeomorphism constaint and obtain the corresponding physical Hamiltonian \cite{Husain:2021ojz,Cipriani:2024nhx}. Quantum gravity effects are then incorporated by introducing suitable modifications to this physical Hamiltonian, yielding an effective quantum Hamiltonian that governs the resulting quantum dynamics. To better understand the problem on the dust shell in these models, let us reinterpret the above approach as follows. Observing that under the dust time gauge, the total diffeomorphism constraint becomes equivalent to that of pure gravity, we begin with the model of pure gravity. By imposing the areal gauge and solving the diffeomorphism constraint, 
the reduced Hamiltonian constraint under the areal gauge is obtained. Incorporating suitable quantum gravity modifications into the reduced Hamiltonian constraint, we then obtain an effective Hamiltonian constraint. To couple a dust field, one can only add the corresponding dust Hamiltonian to the effective Hamiltonian constraint, as there are no remaining constraints in the model. This yields the total effective Hamiltonian constraint. Since the diffeomorphism constraint has already been solved, this total effective Hamiltonian constraint is valid only in gauges where the matter contribution to the diffeomorphism constraint vanishes. The dust time gauge is one such gauge. 
This alternative understanding enables the coupling of additional matter fields to the loop quantum BH models. To achieve this, one simply needs to add the corresponding matter Hamiltonian to the effective Hamiltonian constraint. However, since the contribution of matter to the diffeomorphism constraint is effectively omitted, the resulting total effective Hamiltonian constraint remains valid only under the appropriate gauge conditions, as discussed above. Then, the problem arises when more than two matter fields are coupled: it is not guaranteed that there exists a gauge in which the contributions of all matter fields to the diffeomorphism constraint simultaneously vanish. This indicates a potential issue in the model when a dust shell forms. In particular, neither the PG gauge nor the dust time gauge eliminates the contribution of the dust shell to the diffeomorphism constraint.

As the above discussion illustrates, the presence of a dust shell in the collapsing dust ball model introduces several subtle issues that warrant a more careful analysis. This motivates us to focus on spherically symmetric effective loop quantum gravity coupled to a single dust shell, omitting the collapsing dust ball for simplicity. In light of the aforementioned issues, we need to restore the diffeomorphism constraint in the vacuum model. Furthermore, to address the issue of general covariance, we will treat PG gauge-fixing conditions as additional constraints. In other words, the effective loop quant BH model for the vacuum case will be reformulated as a totally constrained second-class system, with the complete set of constraints including the effective Hamiltonian constraint, the diffeomorphism constraint, and the gauge-fixing conditions. After the reformulation, the dust shell will be coupled, and then its dynamics will be analysis.

In the model with the dust shell coupled, the metrics inside and outside the shell correspond to vacuum solutions with different BH mass parameters. As a result, the only nontrivial dynamical degree of freedom is the radius of the dust shell. From this perspective, to determine the dynamics of the entire system, one can seek an effective action in which the shell radius remains the dynamical variable, while the gravitational fields are treated as fixed. This effective action can be achieved, roughly speaking, by first considering the total action and then substituting the vacuum solutions for the interior and exterior regions of the shell (see, e.g., \cite{Friedman:1997fu,Fiamberti:2007qb} for such formalism  in GR). Consequently, the reduced action becomes a functional of the shell radius and the mass parameters characterizing the vacuum geometries on either side of the shell.  The dynamics of the shell can then be derived by varying this reduced action. This approach has the advantage of bypassing the need to analyze weak solutions to the equations of motion, thereby avoiding ambiguities arising from different choices of variables that may lead to inequivalent weak solutions \cite{Fazzini:2025hsf}.

This paper is organized as follows. In Sec. \ref{sec:two}, the vacuum effective loop quant BH model is reformulated by introducing its action. In Sec. \ref{sec:three}, the model with the dust shell coupled is introduced. In Sec. \ref{sec:four}, the dynamics for the dust shell is derived by calculating the reduced action. In Sec. \ref{sec:EOMforr}, we show some numerical results to the equations of motion for the dust shell. Finally, our approach is concluded and compared with the previous works in Sec. \ref{sec:Comparison}.

\section{Action of the vacuum model}\label{sec:two}
In the original works such as \cite{Kelly:2020uwj,Husain:2022gwp}, the effective theory of loop quantum BH models is formulated within the Hamiltonian framework. To reformulate the model, we translate it into a Lagrangian framework by defining a suitable action. To begin with,  consider the spacetime manifold $\mathcal M_2\times \mathbb S^2$, with $\mathbb S^2$ denoting the 2-sphere and $\mathcal M_2$ being some 2-dimensional manifold. Due to the symmetry, the theory is reduced to dilaton gravity on the 2-manifold  $\mathcal M_2\cong \mathbb R\times\Sigma$ with $\Sigma$ being the spatial manifold, which, for example, could be the positive real line.   The action of the loop quantum BH model for vacuum reads
\begin{equation}\label{eq:actioneff}
\begin{aligned}
S_g=&-\int\dd t\int\dd x\Bigg(\frac{1}{2}K_1\dot E^1+K_2\dot E^2\\
&+N\heff+N^xH_x+\lambda_IF^I\Bigg),
\end{aligned}
\end{equation}
where  the geometrized unit system with $G=c=1$ is chosen, $H_x$ and $\heff$ are given by \cite{Giesel:2023hys}
\begin{equation}\label{eq:hheff}
\begin{aligned}
H_x=&\frac{1}{2}\left(-K_1\partial_xE^1+2E^2\partial_xK_2\right),\\
\heff=&-\frac{E^2}{2 \sqrt{E^1}}-\frac{3 \sqrt{E^1}E^2\sin ^2\left(\frac{\zeta  K_2}{\sqrt{E^1}}\right)}{2 \zeta ^2 }\\
&+\frac{E^2K_2 \sin \left(\frac{2 \zeta  K_2}{\sqrt{E^1}}\right)}{2 \zeta  }-\frac{E^1K_1\sin \left(\frac{2 \zeta  K_2}{\sqrt{E^1}}\right)}{2 \zeta  }\\
&+\frac{(\partial_xE^1)^2}{8 \sqrt{E^1} E^2}+\frac{\sqrt{E^1} \partial_x^2E^1}{2  E^2}-\frac{\sqrt{E^1} \partial_xE^1\partial_xE^2}{2 (E^2)^2},
\end{aligned}
\end{equation}
with $\zeta\sim \sqrt{\hbar}$ being the quantum parameter, and $F_I$ are the gauge fixing condition given by 
\begin{equation}
F_1=E^1-x^2,\quad F_2=E^1-(E^2)^2.
\end{equation}
Obviously, the model with the action $S_g$ is a totally constrained system of the second class, with the effective Hamiltonian constraint $\heff$, the diffeomorphism constraint $H_x$ and the gauge fixing condition $F_I$. As discussed in Sec. \ref{sec:intro}, the gauge fixing terms are introduced due to the covariance issue. 

 The variation of $S_g$ with respect to $N^x$ and $N$ results in the diffeomorphism and Hamiltonian constraints, respectively. The variation with respect to $\lambda_I$ leads to the gauge fixing conditions
\begin{equation}\label{eq:gaugeE}
E^1(x) = x^2, \quad E^2(x) = \sqrt{E^1}(x).
\end{equation}
With these results, one can solve the values of $K_I$ from the constraints $H_x=0=\heff$:
\begin{equation}\label{eq:solveK}
\begin{aligned}
K_2&=\pm \frac{\sqrt{E^1}}{\zeta} \arcsin\left(\frac{1}{2} \frac{\zeta}{\sqrt{E^1}}  \sqrt{\frac{(\partial_xE^1)^2}{ (E^2)^2}+\frac{8 M}{\sqrt{E^1}}-4}\right),\\
K_1&=\frac{2 E^2}{\partial_xE^1}\partial_xK_2.
\end{aligned}
\end{equation}
for some constant $M$. Then, the values of $N$ and $N^x$ can be obtained by the evolution equation $\delta S/\delta K_I=0$, i.e.,
\begin{equation}\label{eq:solvNNx}
\dot E_I=-\frac{\delta}{\delta K^I}\left(\heff[N]+H_x[N^x]\right)
\end{equation}
where $F[g]\equiv \int F(x)g(x)\dd x$. 
Solving Eqs. \eqref{eq:gaugeE}, \eqref{eq:solveK} and \eqref{eq:solvNNx}, we get the lapse function $N$, shift vector $N^x$, and the triad components $E^1$, $E^2$. The line element is obtained by substituting the results into its general form 
\begin{equation}
\dd s^2=-N^2\dd t^2+\frac{(E^2)^2}{E^1}(\dd x^2+N^x\dd t)^2+E^1\dd\Omega^2.
\end{equation}
In this paper, we are concerned with the stationary spacetime metric:
\begin{equation}
\dd s^2=-\dd t^2+\left(\dd x\pm\sqrt{\frac{2M}{x}\left(1-\frac{2\zeta^2M }{x^3}\right)}\dd t\right)^2+x^2\dd\Omega^2,
\end{equation}
where $\dd\Omega^2$ denotes the standard metric on $\mathbb S^2$ (see \cite{Cafaro:2024vrw} for discussion on different types of solutions of the model). 

\section{The model with dust shell coupled}\label{sec:three}
To construct the total action of the model with the dust shell coupled, we start from the Hamiltonian framework for the dust shell. 
As shown in \cite{Hajicek:1997va,Friedman:1997fu,Fiamberti:2007qb}, the phase space of the dust shell contains its radius $\tx$ and the conjugate momentum $\tp$ with the Poisson bracket 
\begin{equation}
\{\tx,\tp\}=1.
\end{equation}
The dust part of the diffeomorphism constraint is 
\begin{equation}
H_x^{\rm sh}=-\tp\delta(x-\tx),
\end{equation}
and the dust Hamiltonian is 
\begin{equation}\label{eq:Hsh}
H^{\rm sh}=\delta(x-\tx)\sqrt{m^2+\frac{E^1}{(E^2)^2}\tp^2},
\end{equation}
where $m$ is the rest mass of the shell. With these formulas, we suggest the following action for the model with dust shell coupled 
\begin{equation}\label{eq:actioneffec}
\begin{aligned}
S=&\int\dd t\Bigg[\tp\dot\tx-\int\dd x\Bigg(\frac{1}{2}K_1\dot E^1+K_2\dot E^2\\
&+NH_{\rm eff}^{\rm tot}+N^xH_x^{\rm tot}\Bigg)+\lambda_I\tilde F_I\Bigg]+S_\partial,
\end{aligned}
\end{equation}
where we define
\begin{equation}\label{eq:constrainttotal}
H_{\rm eff}^{\rm tot}=\heff+H^{\rm sh},\quad H_x^{\rm tot}=H_x+H_x^{\rm sh}, 
\end{equation}
$\tilde F_I$ and $S_\partial$ denotes the gauge-fixing term and the boundary action respectively. Both of them still need to be determined. We will see below that due to the inclusion of the dust shell, the gauge fixing conditions have to be modified in a neighborhood of the dust shell. 

\subsection{Falloff condition and boundary action}
To be more specific, let us focus on a spacetime region covered by a coordinate chart $(t,x,\theta,\phi)$ with $(t,x)\in(t_i,t_f)\times (x_0,\infty)$.  Before introducing the boundary action, it is necessary to specify the falloff conditions of the fields as $x \to x_0$ and $x \to \infty$. As $x \to x_0$ and $x \to \infty$, the system returns to the vacuum case. Therefore, we require consistency with the gauge conditions introduced in Eqs. \eqref{eq:gaugeE}, \eqref{eq:solveK}, and \eqref{eq:solvNNx}. This motivates us to impose the following falloff behavior:
\begin{equation}\label{eq:falloff1}
\begin{aligned}
&E^1 \sim x^2, \quad E^2 \sim x, \\
&N^x \sim \sqrt{\frac{2M_\pm}{x}\left(1 - \frac{2\zeta^2 M_\pm}{x^3}\right)}, \quad N \sim 1, \\
&K_2 \sim -\frac{x}{\zeta} \arcsin\left( \sqrt{\frac{2\zeta^2 M_\pm}{x^3}} \right), \quad K_1 \sim \partial_x K_2,
\end{aligned}
\end{equation}
where $A \sim B$ means that $A$ asymptotically approaches $B$ as $x \to \infty$ or $x \to x_0$. The constants $M_+$ and $M_-$ denote the mass parameters as $x \to \infty$ and $x \to x_0$, respectively. In Eq.~\eqref{eq:falloff1}, we have chosen a specific sign for $K_2$, as our interest lies in modeling BH formation through the collapse of a dust shell. To describe this physical process, we adopt the ingoing PG coordinates, which naturally selects the sign for $K_2$ as in  Eq.~\eqref{eq:falloff1}.  To be consistent with the falloff condition, the sign in Eq.~\eqref{eq:solveK} must also be fixed to the minus one, i.e. $K_I$ is related to $E^I$ by
\begin{equation}\label{eq:solveKinfalloff}
\begin{aligned}
K_2&=-\frac{\sqrt{E^1}}{\zeta} \arcsin\left( \frac{\zeta}{2\sqrt{E^1}}  \sqrt{\frac{(\partial_xE^1)^2}{ (E^2)^2}+\frac{8 M_\pm}{\sqrt{E^1}}-4}\right),\\
K_1&=\frac{2 E^2}{\partial_xE^1}\partial_xK_2.
\end{aligned}
\end{equation}
It is worth noting that the falloff conditions in Eq.~\eqref{eq:falloff1} not only constrain the allowed configurations of the fields, but also associate to each point in phase space two scalars, $M_\pm$. These scalars define two functions $M_\pm$ on the phase space. For example, $M_+$ can be given by explicitly as
\begin{equation}
M_+:(K_I,E^I)\mapsto -\frac{1}{2}\lim_{x\to\infty}x (K_2)^2.
\end{equation}

Now, let us consider the variation of the action. Because of $\delta E^I\big|_{\infty}=0=\delta E^I\big|_{x_0}$, we have
\begin{equation}\label{eq:dS}
\begin{aligned}
\delta S=&-\int\dd t\left(N^xE^2\delta K_2\right)\Big|_{x_0}^\infty+\delta S_\partial+\cdots\\
=&\int\dd t(\delta M_+-\delta M_-)+\delta S_\partial+\cdots,
\end{aligned}
\end{equation}
where the falloff condition \eqref{eq:falloff1} is applied, and the dots denote those terms corresponding to equations of motion. Equation \eqref{eq:dS} suggests the following boundary action
\begin{equation}
S_\partial=-\int\dd t(M_+-M_-).
\end{equation}

\subsection{Smoothness}
Due to the presence of the dust shell, we assume that the fields $E^I$, $K_I$ are smooth everywhere except at the location $\tx$ of the dust shell.  Furthermore, at $\tx$, the fields $E^I$ are required to be continuous as in the classical theory \cite{Friedman:1997fu,Fiamberti:2007qb}. In particular, the continuity of $E^1$ ensures the continuity of the areal radius of the dust shell, while the continuity of $E^2$ guarantees that Eq.~\eqref{eq:Hsh} is well-defined.
Due to this assumption, enforcing the constraints at the shell, i.e., $H_{\rm eff}^{\rm tot}(\tx)=0=H_x^{\rm tot}(\tx)$, yields the following junction conditions:
\begin{equation}\label{eq:equation11}
[\partial_xE^1]=-2 \sqrt{\tp^2+m^2\frac{(\hat E^2)^2}{\hat E^1}}
\end{equation}
and
\begin{equation}\label{eq:equation12}
[K_2]=\frac{\tp}{\hat E^2}
\end{equation}
where $[A]:=\lim_{\epsilon\to 0}A(\tx+\epsilon)-A(\tx-\epsilon)$ and $\hat A:=A(\tx)$. 

\subsection{Gauge fixing}
To be consistent with the falloff conditions, one might consider adopting the same gauge as in  Eq.~\eqref{eq:gaugeE}, as in the vacuum case. However, this gauge fixing condition $F_I$ is not admissible in the presence of the shell, as the junction condition \eqref{eq:equation11} cannot be satisfied under this gauge. Due to this issue, we have to avoid using the gauge in the entire spacetime. Since the junction conditions \eqref{eq:equation11} and \eqref{eq:equation12} 
retain the same form as in the classical theory, we  follows the proposal given in \cite{Friedman:1997fu,Fiamberti:2007qb} and  adopt the following gauge choice:
\begin{equation}\label{eq:regularizedEs}
\begin{aligned}
\tilde F_1(x)=E^1(x)-&\left[x-\frac{l}{\tx} \sqrt{\tp^2+m^2}f\left(\frac{\tx-x}{l}\right)\right]^2,\\
\tilde F_2(x)=E^2(x)-&\sqrt{E^1(x)},
\end{aligned}
\end{equation}
where $l \ll 1$ is a fixed parameter, $f$ is a function obeying the following properties: 
\begin{itemize}
\item[(i)] $f$ is continuous everywhere, and smooth except at $x=0$; 
\item[(ii)] At $x=0$, we have $f'(0^+)=1$ and $f'(0^-)=0$;
\item[(iii)]  $f(x)=0$ for $x\notin (0,1)$. 
\end{itemize}
Due to the properties of $f$, the gauge \eqref{eq:regularizedEs} is compatible with the one in Eq.~\eqref{eq:gaugeE} for $x\in (x_0,\tx-l)\cup (\tx,\infty)$. 
Because of the property $(ii)$, we have
\begin{equation}\label{eq:dErmp}
\begin{aligned}
(\partial_xE^1)(\tx^+)=&2\tx\\
(\partial_xE^1)(\tx^-)=&2\tx+2\sqrt{m^2+\tp^2}.
\end{aligned}
\end{equation}
Thus, the junction condition \eqref{eq:equation11} is satisfied. 
By substituting Eq.~\eqref{eq:dErmp} into Eq.~\eqref{eq:solveKinfalloff}, we find that, in order for the junction condition \eqref{eq:equation12} to be satisfied, the following condition must be imposed:
\begin{equation}\label{eq:dKconseff}
\begin{aligned}
&\arcsin\left(\frac{\zeta}{\tx^2}  \sqrt{\big[\sqrt{m^2+\tp^2}+\tx\big]^2-\left(\tx^2-2M_-\tx\right)}\right)\\
&-\arcsin\left(  \sqrt{\frac{2\zeta^2M_+}{\tx^3}}\right)=\frac{\zeta \tp}{\tx^2}
\end{aligned}
\end{equation}
which determines $\tp$ as a function of $M_\pm$ and $\tx$. One can also check that for large $\tx$ and small $\zeta$, Eq.~\eqref{eq:dKconseff} reduces to $M_+-M_- \approx \sqrt{m^2+\tp^2}$, as it should be. 

Since the gauge-fixing conditions given in Eq.~\eqref{eq:regularizedEs} differ from those in the vacuum case within the region $x \in (\tilde{x} - l, \tilde{x})$, the dynamics in this interval are no longer consistent with those described by the PG coordinates. However, we will see from the analysis below that the dynamics of the dust shell is independent of the choice of $l$. Thus, the final theory will be defined as the one with $l\to 0$.

\section{The reduced action in the gauge choice}\label{sec:four}
Applying Eq.~\eqref{eq:solveKinfalloff}, a straightforward calculation shows
\begin{equation}\label{eq:relationK12}
\begin{aligned}
\frac{\partial K_1}{\partial E^2}=2\left(\frac{\partial K_2}{\partial E^1}-\partial_x\frac{\partial K_2}{\partial(\partial_xE^1)}\right)
\end{aligned}
\end{equation}
Then, let us define $X$ as a function of $E^1$, $\partial_xE^1$ and $E^2$ by
\begin{equation}\label{eq:defineFeff}
\begin{aligned}
X(E^1,\partial_xE^1,E^2)&=2\int\dd E^2 K_2,\\
X(E^1,\partial_xE^1,\frac{1}{2}\partial_xE^1)&=0
\end{aligned}
\end{equation}
where $K_2$ as a function of $E^2$ is given by Eq.~\eqref{eq:solveKinfalloff}. Note that the second equation in \eqref{eq:defineFeff} fixes the integration constant \footnote{We thank Shengzhi Li for the useful discussion here.}. With this definition, $K_1$ and $K_2$ can be expressed in terms of $X$ as follows:
\begin{equation}\label{eq:fineff}
\begin{aligned}
K_2=&\frac{1}{2}\frac{\partial X}{\partial E^2},\\
K_1=&\frac{\partial X}{\partial E^1}-\partial_x\left(\frac{\partial X}{\partial(\partial_xE^1)}\right).
\end{aligned}
\end{equation}
Indeed, the first equation is a direct result of Eq.~\eqref{eq:defineFeff}. For the second one, it can be obtained by substituting Eq.~\eqref{eq:defineFeff} into it and then applying Eq.~\eqref{eq:relationK12}. The detailed derivation is deferred to Appendix~\ref{app:intconstant} to maintain the flow of the main discussion.
Thanks to the function $X$, the value of the action in the gauge choice \eqref{eq:regularizedEs} becomes 
\begin{equation}\label{eq:action}
S=\int \dd t(\,\tp\dot\tx+L_g\,)
\end{equation}
with
\begin{equation}\label{eq:lg0}
\begin{aligned}
L_g=&-M_++M_--\frac{1}{2}\int_{x_0}^\infty\left(K_1\dot E^1+2K_2\dot E^2\right)\dd x\\
=&-M_++M_--\frac{1}{2}\int_{x_0}^\tx\dot X\dd x+\frac{\dot M_- }{2}\int_{0}^\tx\frac{\partial X}{\partial M_-}\dd x\\
&+\frac{1}{2G}\frac{\partial X}{\partial(\partial_xE^1)}\dot E^1\Big|^{\tx^-}_{x_0}\\
=&-M_++M_-+\frac{\dot M_- }{2}\int_{0}^\tx\frac{\partial X}{\partial M_-}\dd x+\mathcal Z.
\end{aligned}
\end{equation}
where the total derivative term $-\frac{1}{2}\frac{\dd}{\dd t}\int_{x_0}^\tx X\dd x$ has been omitted in the final result, and $\mathcal Z$ is
\begin{equation}\label{eq:zresult}
\mathcal Z=\left(\frac{1}{2}X\dot \tx+\frac{1}{2}\frac{\partial X}{\partial(\partial_xE^1)}\dot E^1\right)_{x=\tx^-}
\end{equation}
In this work, we will treat $M_-$ as a datum and, thus, the term proportional to $\dot M_-$ vanishes. As a consequence, $L_g$ reads
\begin{equation}\label{eq:lg0}
\begin{aligned}
L_g=-M_++M_-+\mathcal Z.
\end{aligned}
\end{equation}
The calculation for the expression of $\mathcal Z$ can be found in Appendix \ref{app:A}. With the result therein, we finally have
\begin{equation}\label{eq:Lgfinal}
\begin{aligned}
L_g=&-\tp\dot \tx-M_++M_-\\
&+\dot\tx\frac{\tx^2}{\zeta}\Theta(0,M_-)- \dot\tx\frac{\tx^2 }{\zeta }\Theta(0,M_+)\\
&+\dot \tx\frac{\tx^2\sqrt{\tx}}{\zeta\sqrt{\tx-2M_-}}F\left(\Theta(\tp,M_+),\mu\right)\\
&- \dot\tx\frac{\tx^2\sqrt{\tx}}{\zeta\sqrt{\tx-2M_-}}F\left(\Theta(0,M_-),\mu\right)
\end{aligned}
\end{equation}
where $F$ denotes the elliptic integral of the first kind (see Eq.~\eqref{eq:ellip}), and  $\Theta(\tp,M)$ and $\mu$ are 
\begin{equation}
\begin{aligned}
\Theta(\tp,M)=&\frac{\zeta \tp}{\tx^2}+\arcsin(\sqrt{\frac{2\zeta^2M_+}{\tx^3}})\\
\mu=&-\frac{\tx^3}{\zeta ^2 (\tx-2  M_-)}
\end{aligned}
\end{equation}
Substituting Eq.~\eqref{eq:Lgfinal} into Eq.~\eqref{eq:action}, we ultimately obtain the expression of the action 
\begin{equation}\label{eq:actionfinal}
S=\int\left(\dot\tx \Pi-M_++M_-\right)\dd t,
\end{equation}
where $\Pi$ reads
\begin{equation}\label{eq:Pi}
    \begin{aligned}
    \Pi=&\frac{\tx^2}{\zeta}\Bigg[- \Theta(0,M_+)+\frac{\sqrt{\tx}F\left(\Theta(\tp,M_+),\mu\right)}{\sqrt{\tx-2M_-}}\\
&+\Theta(0,M_-)-\frac{\sqrt{\tx}F\left(\Theta(0,M_-),\mu\right)}{\sqrt{\tx-2M_-}}\Bigg].
    \end{aligned}
\end{equation}
 \section{Equation of motion for $\tx$}\label{sec:EOMforr}
According to Eq.~\eqref{eq:dKconseff}, $\tp$ is a function of $M_+$ and $\tx$. In addition, $M_-$ is a constant as stated below Eq.~\eqref{eq:zresult}. Consequently, the action $S$ is a functional of $M_+(t)$ and $\tx(t)$. 
The Euler-Lagrangian equations read
\begin{equation}\label{eq:ELequation}
\begin{aligned}
    0=&\dot\tx\frac{\partial\Pi}{\partial\tx}-\frac{\dd}{\dd t}\Pi=-\frac{\partial \Pi}{\partial M_+}\dot M_+,\\
    0=&\dot\tx\frac{\partial \Pi}{\partial M_+}-1.
\end{aligned}
\end{equation}
These two equations make $\dot M_+=0$, which is consistent with the fact that the mass parameter $M_+$ in the exterior vacuum region remains constant under dynamic. The second equation in Eq.~\eqref{eq:ELequation} gives
\begin{equation}
\begin{aligned}
1=&-\dot\tx\frac{\tx^2 }{\zeta }\partial_{M_+}\Theta(0,M_+)+\\
&+\dot \tx\frac{\tx^2\sqrt{\tx}}{\zeta\sqrt{\tx-2GM_-}}\partial_{M_+}F\left(\Theta(\tp,M_+),\mu\right)
\end{aligned}
\end{equation}
We have
\begin{equation}\label{eq:EOMforr}
\begin{aligned}
&\partial_{M_+}F\left(\Theta(\tp,M_+),\mu\right)\\
=&\frac{1}{\sqrt{1+\mu\sin^2\left(\Theta(\tp,M_+)\right)}}\partial_{M_+}\Theta(\tp,M_+)
\end{aligned}
\end{equation}
With the calculations shown in Appendix \ref{app:B}, this equation can be simplified into 
\begin{equation}\label{eq:EOMfinaleff1}
\begin{aligned}
\dot\tx= &-\left(1+\frac{2 \zeta \tp\tx}{\sqrt{m^2+\tp^2} \left[2 \zeta \tp-\tx^2 \sin \left(2\Theta(\tp,M_+)\right)\right]}\right)\\
 &\sqrt{\frac{2M_+}{\tx}\left(1-\frac{2 \zeta^2 M_+}{\tx^3}\right)}
\end{aligned}
\end{equation}
It is noted that, in our calculation, the gauge choice results in the PG coordinates for $x\in (x_0,\tx-l)\cup (\tx,\infty)$. In the region $(x_0,\tx-l)$ the BH mass is $M_-$, while for $(\tx,\infty)$, it is $M_+$. This gauge choice indicates that the time coordinate along the shell trajectory is the PG time associated with the exterior mass $M_+$. This discussion clarifies the meaning of $\dot\tx$.

To examine its classical limit, we can expand Eq.~\eqref{eq:EOMfinaleff1} as a power series of $\zeta$. This expansion gives us
\begin{equation}
\dot\tx=\frac{\tp}{\sqrt{m^2+\tp^2}}-\sqrt{\frac{2M_+}{\tx}}+O(\zeta),
\end{equation}
which recovers the equation of motion for the dust shell in classical GR \cite{Hajicek:1997va,Friedman:1997fu}. 

The evolution equation \eqref{eq:EOMfinaleff1}, with $\tp$ related to $\tx$ and $M_+$ via Eq.~\eqref{eq:dKconseff}, can be solved numerically. In the following sections we will focus on the numerical results for the case where $M_-=0$ and $M_+=m$. For $M_-$, it means that the interior region enclosed by the dust shell is described by Minkowski spacetime. 

\subsection{Numerical results for $\tx$}

\begin{figure}
    \centering
    \includegraphics[width=0.5\textwidth]{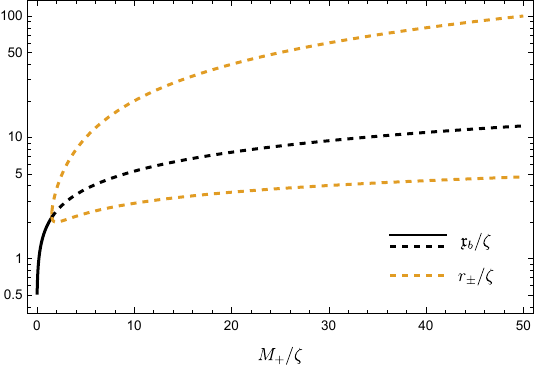}
    \caption{The radius $\tx_b$ at which the dust shell undergoes a bounce. The yellow dashed lines represent the radii of the inner and outer horizons, $r_\pm$. In the black dashed segment, the tangent vector $ \partial_t + \dot{\tx} \partial_x = \partial_t$ to the trajectory of the shell becomes spacelike. Here we consider the cases with $M_-=0$ and $M_+=m$.}
    \label{fig:rb}
\end{figure}

The first interesting scenario is the bouncing behavior of the collapsing dust shell. According to Eqs.~\eqref{eq:EOMfinaleff1} and  \eqref{eq:dKconseff}, by requiring $\dot\tx=0$, we can solve the radius $\tx_b$ at which the dust shell undergoes a bounce. The numerical results are shown in Fig. \ref{fig:rb}, where $\tx_b$ is plotted as a function of $M_+=m$. As shown in Fig. \ref{fig:rb}, when $M_+$ reaches a critical value of approximately $1.554\zeta$, the tangent vector to the trajectory of the dust shell, i.e., $\partial_t + \dot{\tx} \partial_x = \partial_t$, becomes spacelike. This occurs because the bounce takes place within the trapped region of the BH, with recalling that every timelike trajectory in the trapped region inevitably moves toward decreasing radial coordinate.

\begin{figure}[t!]
\centering
\includegraphics[width=0.45\textwidth]{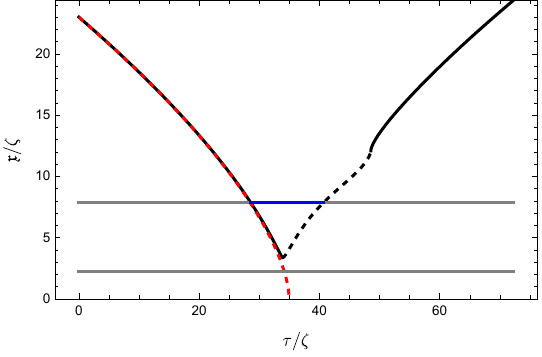}
\caption{Evolution of the dust shell radius $\tx$ as a function of its proper time $\tau$ in the effective loop quantum BH model (black line) and in classical general relativity (red dashed line). The parameters are chosen as $m = 4\zeta=M_+$, and $M_- = 0$. The black dashed segment represents the portion of the trajectory where the shell propagates along a spacelike path. The horizontal lines (gray with blue segment) mark the locations of the inner and outer horizons in the effective spacetime.
}\label{fig:xtau}
\end{figure}

 In Fig. \ref{fig:xtau} we present a numerical result showing $\tx$ as a function of proper time $\tau$, computed for a specific value of $M_+$. As shown in the figure, the dust shell initially undergoes collapse and then transitions into an expanding phase, in contrast to the purely collapsing behavior observed in the classical case.  The black dashed segment indicates that the trajectory becomes spacelike at a certain moment, consistent with the result presented in Fig.~\ref{fig:rb}. Subsequently, the shell follows a spacelike path that crosses the outer horizon, re-emerging from the BH, and eventually transitions back to a timelike trajectory. An important consequence of the spacelike propagation of the dust shell is that the BH horizon becomes finite in duration, as indicated by the blue horizontal segment. In the case considered in Fig. \ref{fig:xtau}, the lifetime of the horizon, i.e., the length of the blue segment, as measured by the proper time of the dust shell is $\Delta \tau\approx 12.158\zeta$.

\begin{figure}[t!]
\centering
\includegraphics[width=0.45\textwidth]{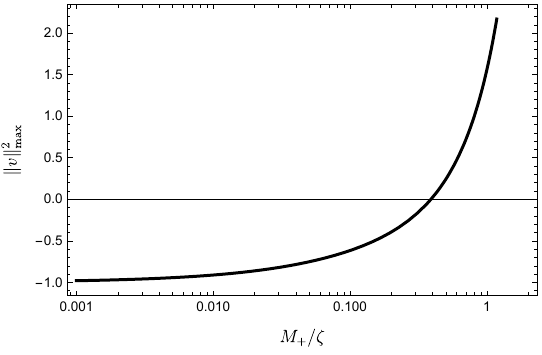}
\caption{The maximal value of $\|v\|^2$ as a function of $M_+$. Here we consider the cases with $M_-=0$ and $M_+=m$.}\label{fig:spacetime}
\end{figure}

It is found that the spacelike segment does not always appear along the trajectory of the dust shell.  In Fig. \ref{fig:spacetime}, we plot the maximal value of $\|v\|^2:=g_{\mu\nu}v^\mu v^\nu$ for $v=\partial_t+\dot\tx\partial_x$ as a function of $M_+/\zeta$. As shown in the figure, when $M_+$ is smaller than a certain threshold, approximately $0.390\zeta$, the quantity $\|v\|^2_{\rm max}$ becomes negative, indicating that the trajectory of the shell remains timelike throughout the entire evolution in such cases. In light of this finding, if Hawking evaporation is taken into account during the collapse process leading to BH formation, it is possible that the issue of spacelike propagation in the dust shell trajectory may no longer arise. In this sense, the result can be interpreted as a partial resolution of the spacelike propagation problem.

\subsection{Determining $N(\tx^-)$ and $N^x(\tx^-)$ and the continuity properties of the metric}
Let us focus on the region with $x\in (\tx-l,\tx)$ to derive the values of $N(\tx^-)$ and $N^x(\tx^-)$ where $\tx^\pm$ is an abbreviation for $\tx\pm \epsilon$. To achieve this, we use Eq.~\eqref{eq:solvNNx} to solve for $N(x)$ and $N^x(x)$ within the interval $x \in (\tilde{x} - l, \tilde{x})$, which gives
\begin{equation}\label{eq:NNxeff}
\begin{aligned}
\dot E^1=&N^x \partial_xE^1 +\frac{E^1 }{\zeta }\sin \left(\frac{2 \zeta  K_2}{\sqrt{E^1}}\right)N,\\
0=&\partial_xN^x-\frac{\partial_x\chi}{\chi}N^x+\frac{\dot E^1}{\partial_xE^1}\frac{\partial_x\chi}{\chi}
\end{aligned}
\end{equation}
where we apply Eq.~\eqref{eq:solveKinfalloff} and the gauge fixing condition $\tilde F_2=0$, and define 
\begin{equation}
\chi=\sqrt{E^1}\sin(\frac{2\zeta K_2}{\sqrt{E^1}}).
\end{equation}
Equation \eqref{eq:NNxeff} can be solved to yield
\begin{equation}\label{eq:intNx}
\begin{aligned}
N^x=&-\chi\int\frac{\dot E^1(y)}{\partial_yE^1(y)}\frac{\partial_y\chi(y)}{\chi(y)^2}\dd y
\end{aligned}
\end{equation}
Substituting $K_2$ given by Eq.~\eqref{eq:solveKinfalloff}, we get the following expression of $\chi$ for $x<\tx$:
\begin{equation}
\begin{aligned}
\chi=&-2\zeta \sqrt{\frac{(\partial_xE^1)^2}{4E^1}+\frac{2 M_-}{\sqrt{E^1}}-1}\times\\
&\sqrt{1-\frac{\zeta ^2}{4E^1}  \left(\frac{(\partial_xE^1)^2}{E^1}+\frac{8 M_-}{\sqrt{E^1}}-4\right)}
\end{aligned}
\end{equation}
We now turn to the more physically interesting case where $M_- = 0$. The analysis can be similarly extended to the general case with $M_- \neq 0$, although the calculations may become more involved.  For $M_-=0$, by inserting the expression of $E^1$ obtained by the gauge fixing condition $\tilde F_1=0$, we find 
\begin{equation}
\begin{aligned}
&\frac{(\partial_xE^1)^2}{4E^1}=\left[1-l\sqrt{m^2+\tp^2} \frac{\partial }{\partial x}\left(\frac{ f\left(\frac{\tx-x}{l}\right)}{x}\right)\right]^2
\end{aligned}
\end{equation}
By the property (ii) of $f$, we have
\begin{equation}\label{eq:dfr}
\frac{\partial }{\partial x}\left(\frac{ f\left(\frac{\tx-x}{l}\right)}{x}\right)\Big|_{x=\tx}=-\frac{1}{l\tx}<0,
\end{equation}
leading to 
\begin{equation}\label{eq:dee1}
\frac{(\partial_xE^1)^2}{4E^1}\Big|_{x=\tx}-1>0.
\end{equation}
In addition, according the property (iii) of $f$, we have
\begin{equation}
\int_{\tx-l}^\tx\dd x\frac{\partial }{\partial x}\left(\frac{ f\left(\frac{\tx-x}{l}\right)}{x}\right)=0
\end{equation}
which together with Eq.~\eqref{eq:dfr} implies that there exist $x_1\in (\tx-l,\tx)$ such that 
\begin{equation}
\frac{\partial }{\partial x}\left(\frac{ f\left(\frac{\tx-x}{l}\right)}{x}\right)\Big|_{x=x_1}>0.
\end{equation}
Consequently, we have 
\begin{equation}\label{eq:dee2}
\frac{(\partial_xE^1)^2}{4E^1}\Big|_{x=x_1}-1<0.
\end{equation}
Combing Eqs.~\eqref{eq:dee1} and \eqref{eq:dee2}, we conclude that there exist $x_0\in (\tx-l,\tx)$ such that 
\begin{equation}
\frac{(\partial_xE^1)^2}{4E^1}\Big|_{x=x_0}-1=0.
\end{equation}
Since $\frac{(\partial_xE^1)^2}{4E^1}$ is proportional to $\chi^2$ with $M_-=0$, it follows that $\chi^2$, when  $M_-=0$, has a root $x_0\in(\tx-l,\tx)$. Since $\chi^2$ appears in the denominator of the integrand in Eq.~\eqref{eq:intNx}, its vanishing at $x_0$ requires careful analysis. If the integral remains finite at $x_0$, then we have
\begin{equation}\label{eq:case1}
N^x(x_0) = 0.
\end{equation}
On the other hand, if the integral diverges at $x_0$, applying the L’H\^opital’s rule leads to
\begin{equation}\label{eq:case2}
N^x(x_0) = \frac{\dot{E}^1(x_0)}{\partial_x E^1(x_0)}.
\end{equation}
The results \eqref{eq:case1} and \eqref{eq:case2} are independent of the lower limit (i.e., the initial condition) of the indefinite integral in Eq.~\eqref{eq:intNx}. This  fact implies that the values of $N^x$ for $x\in (\tx-l,\tx)$ need to be determined by two boundary conditions. In particular, for $x\in (\tx-l,x_0)$, the value is determined by the left boundary condition $N^x(\tx-l)$, and for $x\in (x_0,\tx)$, it is determined by the right boundary condition $N^x(\tx)$. The continuity of the metric at $x=\tx-l$ requires $N^x(\tx-l)=0$. Consequently, $N^x$ can be given by the following piecewise function:
\begin{itemize}
    \item[(1)] For $x\in (\tx-l,x_0)$, we have
    \begin{equation}
        N^x(x)=-\chi(x)\int^x_{\tx-l}\frac{\dot E^1(y)}{\partial_yE^1(y)}\frac{\partial_y\chi(y)}{\chi(y)^2}\dd y;
    \end{equation}
    \item[(2)] For  $x\in (x_0,\tx)$, we have
    \begin{equation}
        N^x(x)=N^x(\tx^-)-\chi(x)\int^x_{\tx-l}\frac{\dot E^1(y)}{\partial_yE^1(y)}\frac{\partial_y\chi(y)}{\chi(y)^2}\dd y,
    \end{equation}
\end{itemize}
 where $N^x(\tx^-)$ is the boundary condition we need to derive in what follows. 

To obtain the boundary condition of $N^x(\tx^-)$, we would impose the continuity of the reduced metric on the world history of the dust shell. This motives us to investigate the squared norm of the tangent vector $\partial_t+\dot\tx\partial_x$ to the trajectory of the dust shell. With noting $\dot E^1(\tx^+)=0$, and
\begin{equation}
[\dot E^1]=-\dot\tx[\partial_xE^1]=2\dot\tx\sqrt{m^2+\tp^2},
\end{equation}
we can simplify the first equation in \eqref{eq:NNxeff} as
\begin{equation}
\begin{aligned}
N^x(\tx^-)= \frac{N(\tx^-) \tx^2 \sin \left(2\Theta(\tp,M^+)\right)-2 \zeta \sqrt{m^2+\tp^2} \dot \tx}{2 \zeta \left(\sqrt{m^2+\tp^2}+\tx)\right)}.
\end{aligned}
\end{equation}
With this result, the squared norm of the tangent vector measured by the interior metric is 
\begin{equation}
\begin{aligned}
&-N(\tx^-)^2+\left(\dot \tx+N^x(\tx^-)\right)\\
=&-\frac{\alpha}{\tx^4(\mathfrak{m}+\tx)^2}\left(N(\tx^-)-\frac{\tx^7 (\tx+\mathfrak{m})^2 \dot\tx \sin \left(2\Theta(\tp,M_+) \right)}{2 \zeta \left(\mathfrak{m}+\tx\right)^2\alpha}\right)^2\\
&+\frac{\tx^6\dot\tx^2}{\alpha}
\end{aligned}
\end{equation}
where $\mathfrak{m}\equiv\sqrt{m^2+\tp^2}$, and 
\begin{equation}
\alpha=\mathfrak{m}^4\zeta^2+4\mathfrak{m}^3\zeta^2\tx+4\mathfrak{m}^2\zeta^2\tx^2+\tx^6.
\end{equation}
Then, if we require the continuity of the reduced metric, i.e., 
\begin{equation}\label{eq:condition1}
-N(\tx^-)^2+\left(\dot \tx+N^x(\tx^-)\right)=-1+\left(\dot\tx+N^x(\tx^+)\right)^2
\end{equation}
we have
\begin{equation}
\begin{aligned}
&-\frac{\alpha}{\tx^4(\mathfrak{m}+\tx)^2}\left(N(\tx^-)-\frac{\tx^7 (\tx+\mathfrak{m})^2 \dot\tx \sin \left(2\Theta(\tp,M_+) \right)}{2 \zeta \left(\mathfrak{m}+\tx\right)^2\alpha}\right)^2\\
=&-\frac{\tx^6\dot\tx^2}{\alpha}-1+\left(\dot\tx+N^x(\tx^+)\right)^2.
\end{aligned}
\end{equation}
When the bounce occurs, i.e., $\dot\tx=0$, we could have $-1+\left(\dot\tx+N^x(\tx^+)\right)^2=-1+N^x(\tx^+)^2>0$, as shown in Fig. \ref{fig:rb}, which will results in complex $N(\tx^-)$. This fact motives us to give up Eq. \eqref{eq:condition1}, and instead we impose 
\begin{equation}\label{eq:condition1}
-N(\tx^-)^2+\left(\dot \tx+N^x(\tx^-)\right)=-\left|-1+\left(\dot\tx+N^x(\tx^+)\right)^2\right|
\end{equation}
which implies the continuity of the length of the trajectory of the dust shell. 
It leads to
\begin{equation}\label{eq:condition12}
\begin{aligned}
&-\frac{\alpha}{\tx^4(\mathfrak{m}+\tx)^2}\left(N(\tx^-)-\frac{\tx^7 (\tx+\mathfrak{m})^2 \dot\tx \sin \left(2\Theta(\tp,M_+) \right)}{2 \zeta \left(\mathfrak{m}+\tx\right)^2\alpha}\right)^2\\
=&-\frac{\tx^6\dot\tx^2}{\alpha}-\left|-1+\left(\dot\tx+N^x(\tx^+)\right)^2\right|,
\end{aligned}
\end{equation}
which ensures that $N(\tx^-)$, and thus $N^x(\tx^-)$, are real. Interestingly, Eq.~\eqref{eq:condition1} implies that the junction surface measure by the interior Minkowski metric is always timelike, although a spacelike segment, as measured by the exterior metric, could occur as shown in Fig. \ref{fig:xtau}. 

\begin{figure}[t!]
\centering
\includegraphics[width=0.45\textwidth]{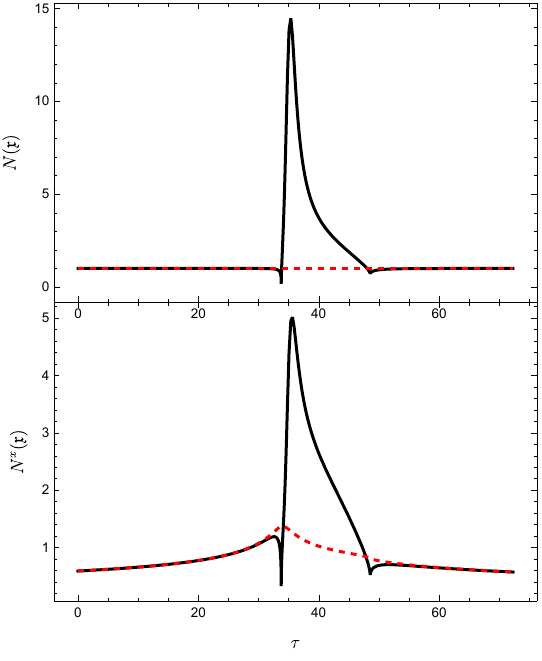}
\caption{The values of $N(\tx)$ and $N^x(\tx)$ along the junction surface as parametrized by its proper time. The black lines plot the values of $N(\tx^-)$ (top panel) and $N^x(\tx^-)$ (bottom panel), while the dashed lines plot $N(\tx^+)=1$ and $N^x(\tx^+)=\sqrt{\frac{2M_+}{\tx}\left(1-\frac{2\zeta^2M_+}{\tx^3}\right)}$ correspondingly. The parameters are $m=4\zeta=M$, $M_-=0$.}\label{fig:NmNx}
\end{figure}

Applying Eq.~\eqref{eq:condition12} and the second equation in Eq.~\eqref{eq:NNxeff}, we numerically calculate 
the values of $N(\tx^-)$ and $N^x(\tx^+)$ with choosing a specific value of $M_+=m$, as shown in Fig. \ref{fig:NmNx}. According to the results, it can be concluded that the metric $g_{\mu\nu}$ is no longer continuous, even though the proper time of the world sheet of the dust shell is continuous. 

\begin{figure}[t!]
\centering
\includegraphics[width=0.45\textwidth]{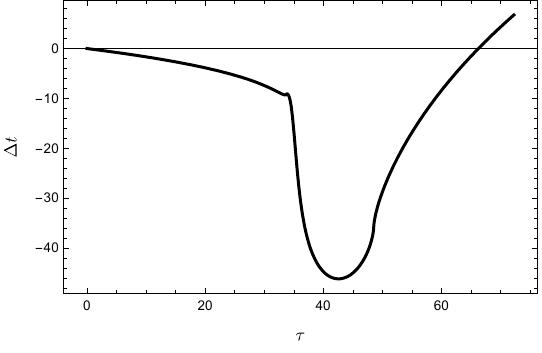}
\caption{The difference between the time coordinates $t$ and $t_-$ as a function of the proper time of the dust shell. Here we choose $m=4\zeta=M_+$. }\label{fig:dttau}
\end{figure}

As discussion below Eq.~\eqref{eq:EOMfinaleff1}, the time coordinate along the shell trajectory is the PG time associated with the exterior mass $M_+$. Since the dynamics of the dust shell is independent of the value of $l$, we can therefore consider the limit $l\to 0$, which will assign to the shell trajectory the 
PG time $t_-$ associated with the interior mass $M_-$. Due to the condition \eqref{eq:condition1}, we have
\begin{equation}
\begin{aligned}
&-\left|-\dd t^2+\left(\dd \tx^2+\sqrt{\frac{2M_+}{\tx}\left(1-\frac{2\zeta^2 M_+}{\tx^3}\right)}\dd t\right)^2\right|\\
=&-\dd t_-^2+\left(\dd \tx^2+\sqrt{\frac{2M_-}{\tx}\left(1-\frac{2\zeta^2 M_-}{\tx^3}\right)}\dd t_-\right)^2,
\end{aligned}
\end{equation}
it is concluded that $t\neq t_-$, in contrary with the presumption used in works such as \cite{Husain:2021ojz,Cipriani:2024nhx}.  A numerical result for the difference $\Delta t=t-t_-$ is presented as a function of the proper time of the dust shell in Fig. \ref{fig:dttau}, where the parameters are chosen as in Fig. \ref{fig:xtau}. 

\begin{figure}[t!]
\centering
\includegraphics[width=0.45\textwidth]{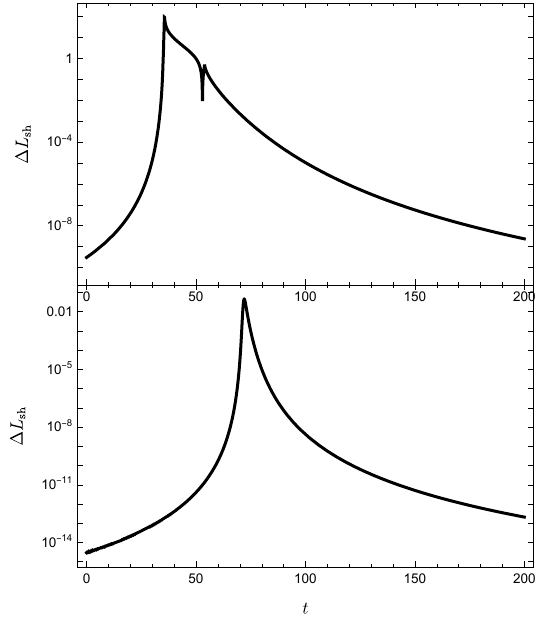}
\caption{The difference between the Lagrangian of the dust shell with its proper time. In the top panel, the parameters are chosen as $M_+=m=4\zeta$ and $M_-=0$. In the bottom panel, we consider a case where there no spacelike segment in the world history of the shell, and the parameters are  $M_+=m=1/3\zeta$ and  $M_-=0$. }\label{fig:dL}
\end{figure}

In this paper, we start with the Hamiltonian formulation of the dust shell to define the action \eqref{eq:actioneffec}, which gives the Lagrangian of dust shell as
\begin{equation}
L_{\rm sh}=\tp\dot\tx-\sqrt{m^2+\tp^2}+\sqrt{\frac{2M_+}{\tx}\left(1-\frac{2\zeta^2 M_+}{\tx^3}\right)}\tp.
\end{equation}
While in the classical theory, the Lagrangian of dust shell is its proper time, i.e., 
\begin{equation}
\tilde L_{\rm sh}=-m\sqrt{\left|1-\left(\dot\tx+\sqrt{\frac{2M_+}{\tx}\left(1-\frac{2\zeta^2 M_+}{\tx^3}\right)}\right)^2\right|}.
\end{equation}
In Fig. \ref{fig:dL}, we present the result for $\Delta L=|L_{\rm sh}-\tilde L_{\rm sh}|$. As demonstrated there, the Lagrangian of the dust shell in the effective loop quantum BH model is no longer equal to its proper time.

\section{Summary and Comparison}\label{sec:Comparison}

 In this paper, we investigate the dynamics of a dust shell coupled to the effective loop quantum BH model. o set the stage, we first reformulate the model by introducing an action that explicitly includes the diffeomorphism constraint, which was originally solved at the outset in the standard formulation, along with appropriate gauge-fixing terms. By adding to the gravitational action the dust shell one, and substituting the vacuum solutions for the interior and exterior regions of the shell, we derive an effective action in which only the shell radius $\tx$ and the exterior mass parameter $M_+$ appear as dynamical variables. By varying this effective action, we finally obtained the evolution equation for $\tx$ which is solved numerically.

 According to the numerical results, the dust shell initially undergoes collapse and then transitions into an expanding phase, in contrast to the purely collapsing behavior seen in the classical case. However, depending on the values of the interior and exterior BH masses $M_\pm$ and the mass of the dust shell $m$, the turning point of the trajectory may lie within the trapped region. In such cases, a spacelike segment, as measured by the exterior metric, appears in the trajectory during the expansion phase. As the shell follows this spacelike path, it eventually crosses the BH horizon, leading to the disappearance of the BH. After remaining spacelike for a brief period, the trajectory eventually transitions back to being timelike. It should be noted that the spacelike segment does not always appear. For example, when choosing $M_+ = m$ and $M_- = 0$, numerical results show that the entire trajectory remains timelike as long as $M_+ < 0.390\zeta$. Moreover, to determined the dynamics of the entire spacetime, the values of $N(\tx^-)$ and $N^x(\tx^-)$ along the junction surface are necessary. It is found that, to have real values for $N(\tx^-)$ and $N^x(\tx^-)$, we cannot require the continuity of the reduced metric along the world history of the dust shell. Instead, the junction condition \eqref{eq:condition1} is imposed which ensures the continuity of the proper time of the dust shell. Interestingly, this junction condition implies that the spacelike segment, as measured by the exterior metric, in the shell trajectory becomes timelike as measured by the interior metric.   
 In this work, the action for the dust shell is constructed starting from its Hamiltonian formulation. We also demonstrate that the corresponding Lagrangian is not simply given by the proper time of the shell.
 
 In the works for instance \cite{Fazzini:2023ova,Cipriani:2024nhx} the evolution of dust shell caused by shell-crossing singularity was also investigated. In those works, the starting point is the full action of a spherically symmetric gravity–dust system. Then, the gauge fixing is performed on the combined system. For our purposes, it is worth noting that the same dynamics can be obtained through an alternative procedure, which can be summarized as follows:
\begin{itemize}
\item[1)] Choose the gauge $E^1=x^2$ and solve the diffeomorphism constraint $H_x=0$ to get $K_1=E^2\partial_xK_2/x$;
\item[2)] Substitute the expression of $E^1$ and $K_1$ into the classical Hamiltonian constraint, and do loop quantization to get the effective Hamiltonian constraint $\tilde H_{\rm eff}$;
\item[3)] Define matter density as $\rho\propto \tilde H_{\rm eff}$;
\item[4)] Find the weak solution to the Hamilton's equation associated with $\tilde H_{\rm eff}$ to get the trajectory of the dust shell. 
\end{itemize}
Although the original procedure presented in the references differs from the one summarized here, the approach described above offers an equivalent alternative interpretation (see Sec.~\ref{sec:intro}). In this procedure, the $\tilde{H}_{\rm eff}$ obtained in step 2) is the same to the expression resulting from substituting the gauge-fixed forms of $E^1$ and $K_1$, as derived in step 1), directly into the effective Hamiltonian constraint $H_{\rm eff}$ given in Eq.~\eqref{eq:hheff} of our paper.
When matter like dust fields and the dust shell is coupled,  its energy density is proportional to the corresponding matter Hamiltonian $H_{\rm matt}$. Therefore, step 3) can be reinterpreted as solving the total Hamiltonian constraint $\tilde{H}_{\rm eff} + H_{\rm matt}$, which is analogous to the first equation in Eq.~\eqref{eq:constrainttotal} in the presence of a dust shell. However, the second equation in Eq.~\eqref{eq:constrainttotal}, representing the total diffeomorphism constraint, has no analogue in this procedure. This is because, in step 1), the diffeomorphism constraint is solved in the absence of matter fields. Consequently, the procedure is only applicable when the contribution of the matter fields to the diffeomorphism constraint vanishes. When dust fields are considered, this requirement is satisfied by adopting the dust-time gauge, in which the dust contribution to the diffeomorphism constraint vanishes. However, when a dust shell forms due to a shell-crossing singularity, the dust-time gauge no longer eliminates the contribution of the dust shell, and the procedure becomes inconsistent in such cases.
Inspired by this discussion, one may consider modifying step 1) by including the diffeomorphism constraint of the dust shell in the equation being imposed. This leads to the relation $K_1=(2E^2\partial_xK_2-H_x^{\rm sh})/(2x)$. It should be noted that the $\delta$-distributional term in $H_x^{\rm sh}$ should be canceled by the corresponding $\delta$-distribution arising from $\partial_x K_2$, ensuring that $K_1$ remains a regular function. Moreover,  gauge choices such as $E^1 = x^2$ and $E^2 = x$ are imposed in those works under discussion. As a result, $\tilde{H}_{\rm eff}$ contains no $\delta$-distributional terms to cancel the one from the energy density of the dust shell, leading to another inconsistency.

 This work develops a new approach for studying the dynamics of a dust shell, which can also be applied to models describing BH formation from the collapse of a dust ball, particularly in scenarios where shell-crossing singularities arise. In addition, the physical interpretation of the spacelike segment in the trajectory of the dust shell remains unclear. Interestingly,  quantitatively similar phenomenon was also found in previous works like \cite{Cipriani:2024nhx,Liu:2025fil}, even though the underlying assumptions and methods differ from those employed in our approach.
 However, it is not yet known whether this feature represents a significant physical phenomenon or is merely an artifact arising from limitations of the model itself, such as the lack of full covariance. Encouragingly, a covariant theory that reproduces the same dynamics in the vacuum case has recently been proposed in \cite{Zhang:2025ccx}. We plan to investigate the dynamics of the dust shell within this covariant framework in future work.
 
\begin{acknowledgements}
C.Z. thanks Francesco Fazzini, Shengzhi Li,  Dongxue Qu and Edward Wilson-Ewing for helpful discussions. This work is supported by the NSFC under Grant No. 12275022, and  “the Fundamental Research Funds for the Central Universities".
H.S. acknowledges COST Action CA23130, supported by COST
(European Cooperation in Science and Technology).

\end{acknowledgements}

\onecolumngrid

\appendix
\section{Derivation of Eq.~\eqref{eq:fineff}}\label{app:intconstant}
The two equations in Eq.~\eqref{eq:defineFeff} can be rewritten equivalently as the following definite integral
\begin{equation}\label{eq:defineFeff0}
\begin{aligned}
X(E^1,\partial_xE^1,E^2)&=2\int_{\frac{\partial_xE^1}{2}}^{E^2} \left(K_2\big|_{E^2=\xi}\right)\dd \xi.
\end{aligned}
\end{equation}
Then, we have
\begin{equation}\label{eq:term1inK1}
\frac{\partial X}{\partial E^1}=2\int_{\frac{\partial_xE^1}{2}}^{E^2} \left(\frac{\partial K_2}{\partial E^1}\Big|_{E^2=\xi}\right)\dd \xi.
\end{equation}
Applying the Leibniz integral rule, we have
\begin{equation}
\begin{aligned}
\frac{\partial X}{\partial(\partial_xE^1)}=2\int_{\frac{\partial_xE^1}{2}}^{E^2} \left(\frac{\partial K_2}{\partial (\partial_xE^1)}\Big|_{E^2=\xi}\right)\dd \xi-\left(K_2\Big|_{E^2=\frac{\partial_xE^1}{2}}\right)
\end{aligned}
\end{equation}
leading to 
\begin{equation}\label{eq:term2inK1}
\begin{aligned}
&\partial_x\left(\frac{\partial X}{\partial(\partial_xE^1)}\right)\\
=&\frac{\partial^2X}{\partial(\partial_xE^1)\partial E^1}\partial_xE^1+\frac{\partial^2X}{\partial(\partial_xE^1)^2}\partial_x^2E^1+\frac{\partial^2X}{\partial(\partial_xE^1)\partial E^2}\partial_xE^2\\
=&2\int_{\frac{\partial_xE^1}{2}}^{E^2} \left(\left(\partial_xE^1\frac{\partial^2 K_2}{\partial (\partial_xE^1)\partial E^1}+\partial_x^2E^1\frac{\partial^2 K_2}{\partial (\partial_xE^1)^2}\right)\Big|_{E^2=\xi}\right)\dd \xi+2\frac{\partial K_2}{\partial(\partial_xE^1)}\partial_xE^2\\
&-\left(\frac{\partial K_2}{\partial (\partial_xE^1)}\Big|_{E^2=\frac{\partial_xE^1}{2}}\right)\partial_x^2E^1-\partial_x\left(K_2\Big|_{E^2=\frac{\partial_xE^1}{2}}\right)
\\
=&2\int_{\frac{\partial_xE^1}{2}}^{E^2} \partial_x\left(\frac{\partial K_2}{\partial (\partial_xE^1)}\right)\Big|_{E^2=\xi}\dd \xi+\left(2\partial_xE^2-\partial_x^2E^1\right)\left(\frac{\partial K_2}{\partial (\partial_xE^1)}\Big|_{E^2=\frac{\partial_xE^1}{2}}\right)-\partial_x\left(K_2\Big|_{E^2=\frac{\partial_xE^1}{2}}\right).
\end{aligned}
\end{equation}
In Eq. \eqref{eq:term2inK1},  to get the last equality, we used 
\begin{equation}
\begin{aligned}
\frac{\partial K_2}{\partial(\partial_xE^1)}-\frac{\partial K_2}{\partial(\partial_xE^1)}\Big|_{E^2=\frac{\partial_xE^1}{2}}=\int_{\frac{\partial_xE^1}{2}}^{E^1}\left(\frac{\partial^2 K_2}{\partial(\partial_xE^1)\partial E^2}\Big|_{E^2=\xi}\right)\dd\xi.
\end{aligned}
\end{equation}
In addition, the integrand $\partial_x\left(\frac{\partial K_2}{\partial (\partial_xE^1)}\right)\Big|_{E^2=\xi}$ in Eq. \eqref{eq:term2inK1}  should be interpreted as
\begin{equation}
\partial_x\left(\frac{\partial K_2}{\partial (\partial_xE^1)}\right)\Big|_{E^2=\xi}=\left(\partial_xE^1\frac{\partial^2 K_2}{\partial (\partial_xE^1)\partial E^1}+\partial_x^2E^1\frac{\partial^2 K_2}{\partial (\partial_xE^1)^2}\right)\Big|_{E^2=\xi}+\left(\frac{\partial^2K_2}{\partial(\partial_xE^1)\partial E^2}\Big|_{E^2=\xi}\right)\partial_xE^2,
\end{equation}
namely, the substitution for $E^2$ is not applied to $\partial_xE^2$, which will be adopted consistently in the following calculations. 
Combining Eqs. \eqref{eq:term1inK1} and \eqref{eq:term2inK1}, we get
\begin{equation}\label{eq:dXeqdXdxE1}
\begin{aligned}
&\frac{\partial X}{\partial E^1}-\partial_x\left(\frac{\partial X}{\partial(\partial_xE^1)}\right)\\
=&2\int_{\frac{\partial_xE^1}{2}}^{E^2} \left(\frac{\partial K_1}{\partial E^2}\Big|_{E^2=\xi}\right)\dd \xi-\left(2\partial_xE^2-\partial_x^2E^1\right)\left(\frac{\partial K_2}{\partial (\partial_xE^1)}\Big|_{E^2=\frac{\partial_xE^1}{2}}\right)+\partial_x\left(K_2\Big|_{E^2=\frac{\partial_xE^1}{2}}\right)\\
=&K_1-K_1\Big|_{E^2=\frac{\partial_xE^1}{2}}-\left(2\partial_xE^2-\partial_x^2E^1\right)\left(\frac{\partial K_2}{\partial (\partial_xE^1)}\Big|_{E^2=\frac{\partial_xE^1}{2}}\right)+\partial_x\left(K_2\Big|_{E^2=\frac{\partial_xE^1}{2}}\right)
\end{aligned}
\end{equation}
where  Eq. \eqref{eq:relationK12} is applied. By Eq. \eqref{eq:solveKinfalloff}, we have
\begin{equation}\label{eq:solveKinfalloff11}
\begin{aligned}
&-K_1\Big|_{E^2=\frac{\partial_xE^1}{2}}+\partial_x\left(K_2\Big|_{E^2=\frac{\partial_xE^1}{2}}\right)=-\left(\partial_xK_2\right)\Big|_{E^2=\frac{\partial_xE^1}{2}}+\left(\frac{\partial K_2}{\partial E^1}\Big|_{E^2=\frac{\partial_xE^1}{2}}\right)\partial_xE^1\\
=&-\partial_xE^2\left(\frac{\partial K_2}{\partial E^2}\right)\Big|_{E^2=\frac{\partial_xE^1}{2}}-\partial_x^2E^1\left(\frac{\partial K_2}{\partial(\partial_xE^1)}\right)\Big|_{E^2=\frac{\partial_xE^1}{2}}.
\end{aligned}
\end{equation}
Substituting this result into Eq. \eqref{eq:dXeqdXdxE1}, we get 
\begin{equation}\label{eq:dXeqdXdxEfinal}
\begin{aligned}
\frac{\partial X}{\partial E^1}-\partial_x\left(\frac{\partial X}{\partial(\partial_xE^1)}\right)=K_1-\partial_xE^2\left(\frac{\partial K_2}{\partial E^2}+2\frac{\partial K_2}{\partial (\partial_xE^1)}\right)\Big|_{E^2=\frac{\partial_xE^1}{2}}
\end{aligned}
\end{equation}
According to Eq. \eqref{eq:solveKinfalloff}, $K_2$ is a function of $E^1$ and $s=\partial_xE^1/E^2$. Thus, we get
\begin{equation}
\begin{aligned}
\left(\frac{\partial K_2}{\partial E^2}+2\frac{\partial K_2}{\partial (\partial_xE^1)}\right)\Big|_{E^2=\frac{\partial_xE^1}{2}}=\left(\frac{\partial K_2}{\partial s}\left(-\frac{\partial_xE^1}{(E^2)^2}+\frac{2}{E^2}\right)\right)\Big|_{E^2=\frac{\partial_xE^1}{2}}=0.
\end{aligned}
\end{equation}
Inserting this result into Eq. \eqref{eq:dXeqdXdxEfinal}, we finally have
\begin{equation}
\frac{\partial X}{\partial E^1}-\partial_x\left(\frac{\partial X}{\partial(\partial_xE^1)}\right)=K_1.
\end{equation}

\section{The expression of $\mathcal  Z$}\label{app:A}
According to the junction condition \eqref{eq:equation11}, we have 
\begin{equation}
[\dot E^1]=-\dot\tx[\partial_xE^1]=2\dot\tx\sqrt{m^2+\tp^2}.
\end{equation}
Thus, we get
\begin{equation}
\dot E^1\big|_{x=\tx^-}=\dot E^1\big|_{x=\tx^+}-2  \sqrt{m^2 + \tp^2}\dot\tx=-2  \sqrt{m^2 + \tp^2}\dot\tx.
\end{equation}
This result coincides with the one calculated by applying Eq.~\eqref{eq:regularizedEs}, i.e., 
\begin{equation}
\dot E^1\big|_{x=\tx^-}=\left(\frac{\partial E^1}{\partial\tx}\dot\tx+\frac{\partial E^1}{\partial\tp}\dot \tp\right)\Big|_{x=\tx^-}=-2  \sqrt{m^2 + \tp^2}\dot\tx.
\end{equation}
Thus, we have
\begin{equation}
\begin{aligned}
\mathcal Z=\frac{1}{2}\left(X-2  \sqrt{m^2+\tp^2}\frac{\partial X}{\partial(\partial_xE^1)}\right)_{x=\tx^-}\dot \tx
\end{aligned}
\end{equation}
According to Eq.~\eqref{eq:defineFeff}, we gave
\begin{equation}\label{eq:definefp}
\begin{aligned}
X=&-2\frac{\sqrt{E^1}}{\zeta} E^2  \arcsin\left(\frac{1}{2} \frac{\zeta}{\sqrt{E^1}}  \sqrt{\frac{(\partial_xE^1)^2}{ (E^2)^2}+\frac{8M}{\sqrt{E^1}}-4}\right)\\
&-2\frac{\sqrt{E^1}}{\zeta}\int \dd E^2 \frac{\zeta  (\partial_xE^1)^2}{(E^2)^2 \sqrt{\frac{(\partial_xE^1)^2}{(E^2)^2}+\frac{8M}{\sqrt{E^1}}-4} \sqrt{4E^1-\zeta ^2 \left(\frac{(\partial_xE1)^2}{(E^2)^2}+\frac{8G M}{\sqrt{E^1}}-4\right)}}+\partial_xE^1f'(E^1)
\end{aligned}
\end{equation}
where $f$ is an arbitrary function. 
Moreover, we have
\begin{equation}\label{eq:dXde}
\begin{aligned}
\frac{\partial X}{\partial(\partial_xE^1)}=-2\frac{\sqrt{E^1}}{\zeta}\int \dd E^2 \frac{\zeta  \partial_xE^1}{(E^2)^2 \sqrt{\frac{(\partial_xE^1)^2}{(E^2)^2}+\frac{8M}{\sqrt{E^1}}-4} \sqrt{4E^1-\zeta ^2 \left(\frac{(\partial_xE1)^2}{(E^2)^2}+\frac{8M}{\sqrt{E^1}}-4\right)}}+f'(E^1)
\end{aligned}
\end{equation}
Combining Eqs.~\eqref{eq:definefp} and \eqref{eq:dXde}, we get 
\begin{equation}\label{eq:coreacton2}
\begin{aligned}
&\left(X-2 \sqrt{m^2+\tp^2}\frac{\partial X}{\partial(\partial_xE^1)}\right)_{x=\tx^-}=-2 \Big(\frac{\tx^2 }{\zeta }\arcsin\left( \sqrt{\frac{2\zeta^2 M_+}{\tx^3}}\right)+\tp\Big)\\
&+2\tx\left(-\frac{\tx}{\zeta}\int^\tx \dd y\frac{\zeta \big[\sqrt{m^2 + \tp^2} + \tx\big]}{ \sqrt{ \big[ \sqrt{m^2 + \tp^2} + \tx\big]^2-\left(1-\frac{2 M_-}{\tx}\right)y^2} \sqrt{\left(\tx^2+\zeta^2\left(1-\frac{2  M_-}{\tx}\right)\right) y^2-\zeta ^2\big[ \sqrt{m^2 + \tp^2} + \tx\big]^2}}+f'(\tx^2)\right)
\end{aligned}
\end{equation}
 where the junction condition \eqref{eq:equation12} and the following consequence of  the junction condition \eqref{eq:equation11} are applied:
\begin{equation}
\Big(\partial_xE^1-2\sqrt{\tp^2+m^2}\Big)_{x=\tx^-}=2\tx.
\end{equation}

For the free function $f$, we need to choose it such that $X=0$ for $\partial_xE^2=2x=2\sqrt{E^1}$ and $E^2=\sqrt{E^1}$. This requirement leads to
\begin{equation}
\begin{aligned}
f'(E^1)=\frac{\sqrt{E^1}}{\zeta} \arcsin\left( \frac{\zeta}{\sqrt{E^1}}  \sqrt{\frac{2 M}{\sqrt{E^1}}}\right)+\int^{\sqrt{E^1}} \dd E^2 \frac{ E^1  }{(E^2)^2 \sqrt{\frac{E^1}{(E^2)^2}+\frac{2  M}{\sqrt{E^1}}-1} \sqrt{2E^1-\zeta ^2 \left(\frac{E^1}{(E^2)^2}+\frac{2 M}{\sqrt{E^1}}-1\right)}}.
\end{aligned}
\end{equation}
Substituting this result into Eq.~\eqref{eq:coreacton2}, we finally can obtain
\begin{equation}\label{eq:Z}
\begin{aligned}
\mathcal Z\dot \tx^{-1}
=&-\tp+\frac{\tx^2}{\zeta} \arcsin\left(  \sqrt{\frac{2 \zeta^2 M_-}{\tx^3}}\right)- \frac{\tx^2 }{\zeta }\arcsin\left( \sqrt{\frac{2\zeta^2M_+}{\tx^3}}\right)\\
&-\int^\tx\frac{\tx^2\big[ \sqrt{m^2 + \tp^2} + \tx\big] \dd y}{ \sqrt{ \big[ \sqrt{m^2 + \tp^2} + \tx\big]^2-\left(1-\frac{2 M_-}{\tx}\right)y^2} \sqrt{\left(\tx^2+\zeta^2\left(1-\frac{2  M_-}{\tx}\right)\right) y^2-\zeta ^2\big[ \sqrt{m^2 + \tp^2} + \tx\big]^2}}\\
&+ \int^{\tx} \dd y \frac{ \tx^3  }{ \sqrt{\tx^2-\left(1-\frac{2  M_-}{\tx}\right)y^2} \sqrt{\left(\tx^2+\zeta^2\left(1-\frac{2M_-}{\tx}\right)\right)y^2-\zeta ^2 \tx^2}}
\end{aligned}
\end{equation}

Let us  introduce $F(\theta,m)$ to denote the elliptic integral of the first kind, i.e., 
\begin{equation}\label{eq:ellip}
F(\theta,m)=\int \frac{1}{\sqrt{1-m\sin^2(\theta)}}\dd\theta.
\end{equation}
We will have
\begin{equation}
\begin{aligned}
&-\int^\tx\frac{\tx^2\big[ \sqrt{m^2 + \tp^2} + \tx\big] \dd y}{ \sqrt{ \big[ \sqrt{m^2 + \tp^2} + \tx\big]^2-\left(1-\frac{2  M_-}{\tx}\right)y^2} \sqrt{\left(\tx^2+\zeta^2\left(1-\frac{2  M_-}{\tx}\right)\right) y^2-\zeta ^2\big[ \sqrt{m^2 + \tp^2} + \tx\big]^2}}\\
=&\frac{\tx^2\sqrt{\tx}}{\zeta\sqrt{\tx-2M_-}}F\left(\frac{\zeta \tp}{\tx^2}+\arcsin(\sqrt{\frac{2\zeta^2M_+}{\tx^3}}),-\frac{\tx^3}{\zeta ^2 (\tx-2  M_-)}\right),
\end{aligned}
\end{equation}
and
\begin{equation}
\begin{aligned}
&\int^{\tx} \dd y \frac{ \tx^3  }{ \sqrt{\tx^2-\left(1-\frac{2  M_-}{\tx}\right)y^2} \sqrt{\left(\tx^2+\zeta^2\left(1-\frac{2M_-}{\tx}\right)\right)y^2-\zeta ^2 \tx^2}}\\
=&-\frac{\tx^2\sqrt{\tx}}{\zeta\sqrt{\tx-2M_-}}F\left(\arcsin(\sqrt{\frac{2\zeta^2M_-}{\tx^3}}),-\frac{\tx^3}{\zeta ^2 (\tx-2  M_-)}\right)
\end{aligned}
\end{equation}
Substituting the results into Eq.~\eqref{eq:Z}, we finally have
\begin{equation}
\begin{aligned}
\mathcal Z=
&-\dot\tx\tp+\dot\tx\frac{\tx^2}{\zeta} \arcsin\left(  \sqrt{\frac{2 \zeta^2M_-}{\tx^3}}\right)- \dot\tx\frac{\tx^2 }{\zeta }\arcsin\left( \sqrt{\frac{2\zeta^2M_+}{\tx^3}}\right)\\
&+\dot \tx\frac{\tx^2\sqrt{\tx}}{\zeta\sqrt{\tx-2M_-}}F\left(\frac{\zeta \tp}{\tx^2}+\arcsin(\sqrt{\frac{2\zeta^2M_+}{\tx^3}}),-\frac{\tx^3}{\zeta ^2 (\tx-2  M_-)}\right)\\
&- \dot\tx\frac{\tx^2\sqrt{\tx}}{\zeta\sqrt{\tx-2M_-}}F\left(\arcsin(\sqrt{\frac{2\zeta^2M_-}{\tx^3}}),-\frac{\tx^3}{\zeta ^2 (\tx-2  M_-)}\right)
\end{aligned}
\end{equation}

\section{Calculations for $\partial_{M_+}\Theta(\tp,M_+)$}\label{app:B}
By definition, we have
\begin{equation}
\partial_{M_+}\Theta(\tp,M_+)=\frac{\zeta }{\tx^2}\partial_{M_+}\tp+\partial_{M_+}\arcsin(\sqrt{\frac{2\zeta^2M_+}{\tx^3}}).
\end{equation}
The second term can be easily computed as
\begin{equation}\label{eq:dmas}
\partial_{M_+}\arcsin(\sqrt{\frac{2\zeta^2M_+}{\tx^3}})=\frac{\sqrt{\zeta ^2 M_+}}{\sqrt{2} M_+ \sqrt{\tx^3-2 \zeta ^2 M_+}}.
\end{equation}
To calculate the first term, we need to differentiate the both sides of Eq.~\eqref{eq:dKconseff} to get 
\begin{equation}\label{eq:dmp}
\begin{aligned}
&\frac{1}{\partial_{M_+}\tp}=
&-\sqrt{\frac{2 M_+}{\tx}}  \sqrt{1-\frac{2 \zeta ^2  M_+}{\tx^3}}\Bigg(1-\frac{2\tp \zeta\left( \sqrt{m^2+\tp^2}+\tx\right)}{\tx^2\sqrt{m^2+\tp^2} \sin(2\Theta(\tp,M_+))}\Bigg),
\end{aligned}
\end{equation}
where we used the following result derived from  Eq.~\eqref{eq:dKconseff}:
\begin{equation}
 \sqrt{\big[\sqrt{m^2+\tp^2}+\tx\big]^2-\left(\tx^2-2M_-\tx\right)}=\frac{\tx^2} {\zeta}\sin( \Theta(\tp,M_+)).
\end{equation}
Combining Eqs.~\eqref{eq:dmas} and \eqref{eq:dmp}, we have
\begin{equation}
\begin{aligned}
\partial_{M_+}\Theta(\tp,M_+)=-\frac{2\tp \zeta\left(\sqrt{m^2+\tp^2}+\tx\right)\frac{\zeta/\tx}{\sqrt{2M_+/\tx} \sqrt{1-2 \zeta ^2 M_+/\tx^3}}}{\tx^2\sqrt{m^2+\tp^2} \sin(2\Theta(\tp,M_+))-2\tp \zeta\left( \sqrt{m^2+\tp^2}+\tx\right)}
\end{aligned}
\end{equation}

Substituting all of this results into Eq.~\eqref{eq:EOMforr}, we can simplify it into
\begin{equation}\label{eq:EOMfinaleff}
\begin{aligned}
&\sqrt{2M_+/\tx} \sqrt{1-2 \zeta ^2  M_+/\tx^3}\\
=&-\dot\tx\Bigg\{1+\frac{1}{\sqrt{\mu^{-1}+\sin^2\left(\Theta(\tp,M_+)\right)}}\frac{2\tp \zeta^2\left( \sqrt{m^2+\tp^2}+\tx\right)/\tx}{\tx^2\sqrt{m^2+\tp^2} \sin(2\Theta(\tp,M_+))-2\tp \zeta\left( \sqrt{m^2+\tp^2}+\tx\right)}\Bigg\}
\end{aligned}
\end{equation}
Next, we used Eq.~\eqref{eq:dKconseff}  again to get 
\begin{equation}
\begin{aligned}
&\sqrt{\mu^{-1}+\sin^2\left(\Theta(\tp,M_+)\right)}=\frac{\zeta}{\tx^2} \Big[\sqrt{m^2+\tp^2}+\tx\Big].
\end{aligned}
\end{equation}
Thus, Eq.~\eqref{eq:EOMfinaleff} can be further simplified to get the final result \eqref{eq:EOMfinaleff1}.


\begin{thebibliography}{46}%
\makeatletter
\providecommand \@ifxundefined [1]{%
 \@ifx{#1\undefined}
}%
\providecommand \@ifnum [1]{%
 \ifnum #1\expandafter \@firstoftwo
 \else \expandafter \@secondoftwo
 \fi
}%
\providecommand \@ifx [1]{%
 \ifx #1\expandafter \@firstoftwo
 \else \expandafter \@secondoftwo
 \fi
}%
\providecommand \natexlab [1]{#1}%
\providecommand \enquote  [1]{``#1''}%
\providecommand \bibnamefont  [1]{#1}%
\providecommand \bibfnamefont [1]{#1}%
\providecommand \citenamefont [1]{#1}%
\providecommand \href@noop [0]{\@secondoftwo}%
\providecommand \href [0]{\begingroup \@sanitize@url \@href}%
\providecommand \@href[1]{\@@startlink{#1}\@@href}%
\providecommand \@@href[1]{\endgroup#1\@@endlink}%
\providecommand \@sanitize@url [0]{\catcode `\\12\catcode `\$12\catcode
  `\&12\catcode `\#12\catcode `\^12\catcode `\_12\catcode `\%12\relax}%
\providecommand \@@startlink[1]{}%
\providecommand \@@endlink[0]{}%
\providecommand \url  [0]{\begingroup\@sanitize@url \@url }%
\providecommand \@url [1]{\endgroup\@href {#1}{\urlprefix }}%
\providecommand \urlprefix  [0]{URL }%
\providecommand \Eprint [0]{\href }%
\providecommand \doibase [0]{http://dx.doi.org/}%
\providecommand \selectlanguage [0]{\@gobble}%
\providecommand \bibinfo  [0]{\@secondoftwo}%
\providecommand \bibfield  [0]{\@secondoftwo}%
\providecommand \translation [1]{[#1]}%
\providecommand \BibitemOpen [0]{}%
\providecommand \bibitemStop [0]{}%
\providecommand \bibitemNoStop [0]{.\EOS\space}%
\providecommand \EOS [0]{\spacefactor3000\relax}%
\providecommand \BibitemShut  [1]{\csname bibitem#1\endcsname}%
\let\auto@bib@innerbib\@empty
\bibitem [{\citenamefont {Ashtekar}\ and\ \citenamefont
  {Lewandowski}(2004)}]{Ashtekar:2004eh}%
  \BibitemOpen
  \bibfield  {author} {\bibinfo {author} {\bibfnamefont {Abhay}\ \bibnamefont
  {Ashtekar}}\ and\ \bibinfo {author} {\bibfnamefont {Jerzy}\ \bibnamefont
  {Lewandowski}},\ }\bibfield  {title} {\enquote {\bibinfo {title} {{Background
  independent quantum gravity: A Status report}},}\ }\href {\doibase
  10.1088/0264-9381/21/15/R01} {\bibfield  {journal} {\bibinfo  {journal}
  {Class. Quant. Grav.}\ }\textbf {\bibinfo {volume} {21}},\ \bibinfo {pages}
  {R53} (\bibinfo {year} {2004})},\ \Eprint
  {http://arxiv.org/abs/gr-qc/0404018} {arXiv:gr-qc/0404018} \BibitemShut
  {NoStop}%
\bibitem [{\citenamefont {Rovelli}(2004)}]{Rovelli:2004tv}%
  \BibitemOpen
  \bibfield  {author} {\bibinfo {author} {\bibfnamefont {Carlo}\ \bibnamefont
  {Rovelli}},\ }\href {\doibase 10.1017/CBO9780511755804} {\emph {\bibinfo
  {title} {{Quantum gravity}}}},\ Cambridge Monographs on Mathematical Physics\
  (\bibinfo  {publisher} {Univ. Pr.},\ \bibinfo {address} {Cambridge, UK},\
  \bibinfo {year} {2004})\BibitemShut {NoStop}%
\bibitem [{\citenamefont {Han}\ \emph {et~al.}(2007)\citenamefont {Han},
  \citenamefont {Huang},\ and\ \citenamefont {Ma}}]{Han:2005km}%
  \BibitemOpen
  \bibfield  {author} {\bibinfo {author} {\bibfnamefont {Muxin}\ \bibnamefont
  {Han}}, \bibinfo {author} {\bibfnamefont {Weiming}\ \bibnamefont {Huang}}, \
  and\ \bibinfo {author} {\bibfnamefont {Yongge}\ \bibnamefont {Ma}},\
  }\bibfield  {title} {\enquote {\bibinfo {title} {{Fundamental structure of
  loop quantum gravity}},}\ }\href {\doibase 10.1142/S0218271807010894}
  {\bibfield  {journal} {\bibinfo  {journal} {Int. J. Mod. Phys. D}\ }\textbf
  {\bibinfo {volume} {16}},\ \bibinfo {pages} {1397--1474} (\bibinfo {year}
  {2007})},\ \Eprint {http://arxiv.org/abs/gr-qc/0509064} {arXiv:gr-qc/0509064}
  \BibitemShut {NoStop}%
\bibitem [{\citenamefont {Thiemann}(2007)}]{Thiemann:2007pyv}%
  \BibitemOpen
  \bibfield  {author} {\bibinfo {author} {\bibfnamefont {Thomas}\ \bibnamefont
  {Thiemann}},\ }\href {\doibase 10.1017/CBO9780511755682} {\emph {\bibinfo
  {title} {{Modern Canonical Quantum General Relativity}}}},\ Cambridge
  Monographs on Mathematical Physics\ (\bibinfo  {publisher} {Cambridge
  University Press},\ \bibinfo {year} {2007})\BibitemShut {NoStop}%
\bibitem [{\citenamefont {Modesto}(2010)}]{Modesto:2008im}%
  \BibitemOpen
  \bibfield  {author} {\bibinfo {author} {\bibfnamefont {Leonardo}\
  \bibnamefont {Modesto}},\ }\bibfield  {title} {\enquote {\bibinfo {title}
  {{Semiclassical loop quantum black hole}},}\ }\href {\doibase
  10.1007/s10773-010-0346-x} {\bibfield  {journal} {\bibinfo  {journal} {Int.
  J. Theor. Phys.}\ }\textbf {\bibinfo {volume} {49}},\ \bibinfo {pages}
  {1649--1683} (\bibinfo {year} {2010})},\ \Eprint
  {http://arxiv.org/abs/0811.2196} {arXiv:0811.2196 [gr-qc]} \BibitemShut
  {NoStop}%
\bibitem [{\citenamefont {Gambini}\ and\ \citenamefont
  {Pullin}(2008)}]{Gambini:2008dy}%
  \BibitemOpen
  \bibfield  {author} {\bibinfo {author} {\bibfnamefont {Rodolfo}\ \bibnamefont
  {Gambini}}\ and\ \bibinfo {author} {\bibfnamefont {Jorge}\ \bibnamefont
  {Pullin}},\ }\bibfield  {title} {\enquote {\bibinfo {title} {{Black holes in
  loop quantum gravity: The Complete space-time}},}\ }\href {\doibase
  10.1103/PhysRevLett.101.161301} {\bibfield  {journal} {\bibinfo  {journal}
  {Phys. Rev. Lett.}\ }\textbf {\bibinfo {volume} {101}},\ \bibinfo {pages}
  {161301} (\bibinfo {year} {2008})},\ \Eprint {http://arxiv.org/abs/0805.1187}
  {arXiv:0805.1187 [gr-qc]} \BibitemShut {NoStop}%
\bibitem [{\citenamefont {Corichi}\ and\ \citenamefont
  {Singh}(2016)}]{Corichi:2015xia}%
  \BibitemOpen
  \bibfield  {author} {\bibinfo {author} {\bibfnamefont {Alejandro}\
  \bibnamefont {Corichi}}\ and\ \bibinfo {author} {\bibfnamefont {Parampreet}\
  \bibnamefont {Singh}},\ }\bibfield  {title} {\enquote {\bibinfo {title}
  {{Loop quantization of the Schwarzschild interior revisited}},}\ }\href
  {\doibase 10.1088/0264-9381/33/5/055006} {\bibfield  {journal} {\bibinfo
  {journal} {Class. Quant. Grav.}\ }\textbf {\bibinfo {volume} {33}},\ \bibinfo
  {pages} {055006} (\bibinfo {year} {2016})},\ \Eprint
  {http://arxiv.org/abs/1506.08015} {arXiv:1506.08015 [gr-qc]} \BibitemShut
  {NoStop}%
\bibitem [{\citenamefont {Ben~Achour}\ \emph {et~al.}(2017)\citenamefont
  {Ben~Achour}, \citenamefont {Brahma},\ and\ \citenamefont
  {Marciano}}]{BenAchour:2016brs}%
  \BibitemOpen
  \bibfield  {author} {\bibinfo {author} {\bibfnamefont {Jibril}\ \bibnamefont
  {Ben~Achour}}, \bibinfo {author} {\bibfnamefont {Suddhasattwa}\ \bibnamefont
  {Brahma}}, \ and\ \bibinfo {author} {\bibfnamefont {Antonino}\ \bibnamefont
  {Marciano}},\ }\bibfield  {title} {\enquote {\bibinfo {title} {{Spherically
  symmetric sector of self dual Ashtekar gravity coupled to matter:
  Anomaly-free algebra of constraints with holonomy corrections}},}\ }\href
  {\doibase 10.1103/PhysRevD.96.026002} {\bibfield  {journal} {\bibinfo
  {journal} {Phys. Rev. D}\ }\textbf {\bibinfo {volume} {96}},\ \bibinfo
  {pages} {026002} (\bibinfo {year} {2017})},\ \Eprint
  {http://arxiv.org/abs/1608.07314} {arXiv:1608.07314 [gr-qc]} \BibitemShut
  {NoStop}%
\bibitem [{\citenamefont {Ashtekar}\ \emph {et~al.}(2018)\citenamefont
  {Ashtekar}, \citenamefont {Olmedo},\ and\ \citenamefont
  {Singh}}]{Ashtekar:2018lag}%
  \BibitemOpen
  \bibfield  {author} {\bibinfo {author} {\bibfnamefont {Abhay}\ \bibnamefont
  {Ashtekar}}, \bibinfo {author} {\bibfnamefont {Javier}\ \bibnamefont
  {Olmedo}}, \ and\ \bibinfo {author} {\bibfnamefont {Parampreet}\ \bibnamefont
  {Singh}},\ }\bibfield  {title} {\enquote {\bibinfo {title} {{Quantum
  Transfiguration of Kruskal Black Holes}},}\ }\href {\doibase
  10.1103/PhysRevLett.121.241301} {\bibfield  {journal} {\bibinfo  {journal}
  {Phys. Rev. Lett.}\ }\textbf {\bibinfo {volume} {121}},\ \bibinfo {pages}
  {241301} (\bibinfo {year} {2018})},\ \Eprint
  {http://arxiv.org/abs/1806.00648} {arXiv:1806.00648 [gr-qc]} \BibitemShut
  {NoStop}%
\bibitem [{\citenamefont {Zhang}\ \emph {et~al.}(2020)\citenamefont {Zhang},
  \citenamefont {Ma}, \citenamefont {Song},\ and\ \citenamefont
  {Zhang}}]{Zhang:2020qxw}%
  \BibitemOpen
  \bibfield  {author} {\bibinfo {author} {\bibfnamefont {Cong}\ \bibnamefont
  {Zhang}}, \bibinfo {author} {\bibfnamefont {Yongge}\ \bibnamefont {Ma}},
  \bibinfo {author} {\bibfnamefont {Shupeng}\ \bibnamefont {Song}}, \ and\
  \bibinfo {author} {\bibfnamefont {Xiangdong}\ \bibnamefont {Zhang}},\
  }\bibfield  {title} {\enquote {\bibinfo {title} {{Loop quantum Schwarzschild
  interior and black hole remnant}},}\ }\href {\doibase
  10.1103/PhysRevD.102.041502} {\bibfield  {journal} {\bibinfo  {journal}
  {Phys. Rev. D}\ }\textbf {\bibinfo {volume} {102}},\ \bibinfo {pages}
  {041502(R)} (\bibinfo {year} {2020})},\ \Eprint
  {http://arxiv.org/abs/2006.08313} {arXiv:2006.08313 [gr-qc]} \BibitemShut
  {NoStop}%
\bibitem [{\citenamefont {Kelly}\ \emph {et~al.}(2020)\citenamefont {Kelly},
  \citenamefont {Santacruz},\ and\ \citenamefont
  {Wilson-Ewing}}]{Kelly:2020uwj}%
  \BibitemOpen
  \bibfield  {author} {\bibinfo {author} {\bibfnamefont {Jarod~George}\
  \bibnamefont {Kelly}}, \bibinfo {author} {\bibfnamefont {Robert}\
  \bibnamefont {Santacruz}}, \ and\ \bibinfo {author} {\bibfnamefont {Edward}\
  \bibnamefont {Wilson-Ewing}},\ }\bibfield  {title} {\enquote {\bibinfo
  {title} {{Effective loop quantum gravity framework for vacuum spherically
  symmetric spacetimes}},}\ }\href {\doibase 10.1103/PhysRevD.102.106024}
  {\bibfield  {journal} {\bibinfo  {journal} {Phys. Rev. D}\ }\textbf {\bibinfo
  {volume} {102}},\ \bibinfo {pages} {106024} (\bibinfo {year} {2020})},\
  \Eprint {http://arxiv.org/abs/2006.09302} {arXiv:2006.09302 [gr-qc]}
  \BibitemShut {NoStop}%
\bibitem [{\citenamefont {Zhang}\ \emph {et~al.}(2022)\citenamefont {Zhang},
  \citenamefont {Ma}, \citenamefont {Song},\ and\ \citenamefont
  {Zhang}}]{Zhang:2021wex}%
  \BibitemOpen
  \bibfield  {author} {\bibinfo {author} {\bibfnamefont {Cong}\ \bibnamefont
  {Zhang}}, \bibinfo {author} {\bibfnamefont {Yongge}\ \bibnamefont {Ma}},
  \bibinfo {author} {\bibfnamefont {Shupeng}\ \bibnamefont {Song}}, \ and\
  \bibinfo {author} {\bibfnamefont {Xiangdong}\ \bibnamefont {Zhang}},\
  }\bibfield  {title} {\enquote {\bibinfo {title} {{Loop quantum deparametrized
  Schwarzschild interior and discrete black hole mass}},}\ }\href {\doibase
  10.1103/PhysRevD.105.024069} {\bibfield  {journal} {\bibinfo  {journal}
  {Phys. Rev. D}\ }\textbf {\bibinfo {volume} {105}},\ \bibinfo {pages}
  {024069} (\bibinfo {year} {2022})},\ \Eprint
  {http://arxiv.org/abs/2107.10579} {arXiv:2107.10579 [gr-qc]} \BibitemShut
  {NoStop}%
\bibitem [{\citenamefont {Husain}\ \emph
  {et~al.}(2022{\natexlab{a}})\citenamefont {Husain}, \citenamefont {Kelly},
  \citenamefont {Santacruz},\ and\ \citenamefont
  {Wilson-Ewing}}]{Husain:2021ojz}%
  \BibitemOpen
  \bibfield  {author} {\bibinfo {author} {\bibfnamefont {Viqar}\ \bibnamefont
  {Husain}}, \bibinfo {author} {\bibfnamefont {Jarod~George}\ \bibnamefont
  {Kelly}}, \bibinfo {author} {\bibfnamefont {Robert}\ \bibnamefont
  {Santacruz}}, \ and\ \bibinfo {author} {\bibfnamefont {Edward}\ \bibnamefont
  {Wilson-Ewing}},\ }\bibfield  {title} {\enquote {\bibinfo {title} {{Quantum
  Gravity of Dust Collapse: Shock Waves from Black Holes}},}\ }\href {\doibase
  10.1103/PhysRevLett.128.121301} {\bibfield  {journal} {\bibinfo  {journal}
  {Phys. Rev. Lett.}\ }\textbf {\bibinfo {volume} {128}},\ \bibinfo {pages}
  {121301} (\bibinfo {year} {2022}{\natexlab{a}})},\ \Eprint
  {http://arxiv.org/abs/2109.08667} {arXiv:2109.08667 [gr-qc]} \BibitemShut
  {NoStop}%
\bibitem [{\citenamefont {Alonso-Bardaji}\ \emph
  {et~al.}(2022{\natexlab{a}})\citenamefont {Alonso-Bardaji}, \citenamefont
  {Brizuela},\ and\ \citenamefont {Vera}}]{Alonso-Bardaji:2021yls}%
  \BibitemOpen
  \bibfield  {author} {\bibinfo {author} {\bibfnamefont {Asier}\ \bibnamefont
  {Alonso-Bardaji}}, \bibinfo {author} {\bibfnamefont {David}\ \bibnamefont
  {Brizuela}}, \ and\ \bibinfo {author} {\bibfnamefont {Ra\"ul}\ \bibnamefont
  {Vera}},\ }\bibfield  {title} {\enquote {\bibinfo {title} {{An effective
  model for the quantum Schwarzschild black hole}},}\ }\href {\doibase
  10.1016/j.physletb.2022.137075} {\bibfield  {journal} {\bibinfo  {journal}
  {Phys. Lett. B}\ }\textbf {\bibinfo {volume} {829}},\ \bibinfo {pages}
  {137075} (\bibinfo {year} {2022}{\natexlab{a}})},\ \Eprint
  {http://arxiv.org/abs/2112.12110} {arXiv:2112.12110 [gr-qc]} \BibitemShut
  {NoStop}%
\bibitem [{\citenamefont {Zhang}(2021)}]{Zhang:2021xoa}%
  \BibitemOpen
  \bibfield  {author} {\bibinfo {author} {\bibfnamefont {Cong}\ \bibnamefont
  {Zhang}},\ }\bibfield  {title} {\enquote {\bibinfo {title} {{Reduced phase
  space quantization of black holes: Path integrals and effective dynamics}},}\
  }\href {\doibase 10.1103/PhysRevD.104.126003} {\bibfield  {journal} {\bibinfo
   {journal} {Phys. Rev. D}\ }\textbf {\bibinfo {volume} {104}},\ \bibinfo
  {pages} {126003} (\bibinfo {year} {2021})},\ \Eprint
  {http://arxiv.org/abs/2106.08202} {arXiv:2106.08202 [gr-qc]} \BibitemShut
  {NoStop}%
\bibitem [{\citenamefont {M\"unch}(2021)}]{Munch:2021oqn}%
  \BibitemOpen
  \bibfield  {author} {\bibinfo {author} {\bibfnamefont {Johannes}\
  \bibnamefont {M\"unch}},\ }\bibfield  {title} {\enquote {\bibinfo {title}
  {{Causal structure of a recent loop quantum gravity black hole collapse
  model}},}\ }\href {\doibase 10.1103/PhysRevD.104.046019} {\bibfield
  {journal} {\bibinfo  {journal} {Phys. Rev. D}\ }\textbf {\bibinfo {volume}
  {104}},\ \bibinfo {pages} {046019} (\bibinfo {year} {2021})},\ \Eprint
  {http://arxiv.org/abs/2103.17112} {arXiv:2103.17112 [gr-qc]} \BibitemShut
  {NoStop}%
\bibitem [{\citenamefont {Gambini}\ \emph {et~al.}(2022)\citenamefont
  {Gambini}, \citenamefont {Olmedo},\ and\ \citenamefont
  {Pullin}}]{Gambini:2022dec}%
  \BibitemOpen
  \bibfield  {author} {\bibinfo {author} {\bibfnamefont {Rodolfo}\ \bibnamefont
  {Gambini}}, \bibinfo {author} {\bibfnamefont {Javier}\ \bibnamefont
  {Olmedo}}, \ and\ \bibinfo {author} {\bibfnamefont {Jorge}\ \bibnamefont
  {Pullin}},\ }\bibfield  {title} {\enquote {\bibinfo {title} {{Towards a
  quantum notion of covariance in spherically symmetric loop quantum
  gravity}},}\ }\href {\doibase 10.1103/PhysRevD.105.026017} {\bibfield
  {journal} {\bibinfo  {journal} {Phys. Rev. D}\ }\textbf {\bibinfo {volume}
  {105}},\ \bibinfo {pages} {026017} (\bibinfo {year} {2022})},\ \Eprint
  {http://arxiv.org/abs/2201.01616} {arXiv:2201.01616 [gr-qc]} \BibitemShut
  {NoStop}%
\bibitem [{\citenamefont {Husain}\ \emph
  {et~al.}(2022{\natexlab{b}})\citenamefont {Husain}, \citenamefont {Kelly},
  \citenamefont {Santacruz},\ and\ \citenamefont
  {Wilson-Ewing}}]{Husain:2022gwp}%
  \BibitemOpen
  \bibfield  {author} {\bibinfo {author} {\bibfnamefont {Viqar}\ \bibnamefont
  {Husain}}, \bibinfo {author} {\bibfnamefont {Jarod~George}\ \bibnamefont
  {Kelly}}, \bibinfo {author} {\bibfnamefont {Robert}\ \bibnamefont
  {Santacruz}}, \ and\ \bibinfo {author} {\bibfnamefont {Edward}\ \bibnamefont
  {Wilson-Ewing}},\ }\bibfield  {title} {\enquote {\bibinfo {title} {{Fate of
  quantum black holes}},}\ }\href {\doibase 10.1103/PhysRevD.106.024014}
  {\bibfield  {journal} {\bibinfo  {journal} {Phys. Rev. D}\ }\textbf {\bibinfo
  {volume} {106}},\ \bibinfo {pages} {024014} (\bibinfo {year}
  {2022}{\natexlab{b}})},\ \Eprint {http://arxiv.org/abs/2203.04238}
  {arXiv:2203.04238 [gr-qc]} \BibitemShut {NoStop}%
\bibitem [{\citenamefont {Alonso-Bardaji}\ \emph
  {et~al.}(2022{\natexlab{b}})\citenamefont {Alonso-Bardaji}, \citenamefont
  {Brizuela},\ and\ \citenamefont {Vera}}]{Alonso-Bardaji:2022ear}%
  \BibitemOpen
  \bibfield  {author} {\bibinfo {author} {\bibfnamefont {Asier}\ \bibnamefont
  {Alonso-Bardaji}}, \bibinfo {author} {\bibfnamefont {David}\ \bibnamefont
  {Brizuela}}, \ and\ \bibinfo {author} {\bibfnamefont {Ra\"ul}\ \bibnamefont
  {Vera}},\ }\bibfield  {title} {\enquote {\bibinfo {title} {{Nonsingular
  spherically symmetric black-hole model with holonomy corrections}},}\ }\href
  {\doibase 10.1103/PhysRevD.106.024035} {\bibfield  {journal} {\bibinfo
  {journal} {Phys. Rev. D}\ }\textbf {\bibinfo {volume} {106}},\ \bibinfo
  {pages} {024035} (\bibinfo {year} {2022}{\natexlab{b}})},\ \Eprint
  {http://arxiv.org/abs/2205.02098} {arXiv:2205.02098 [gr-qc]} \BibitemShut
  {NoStop}%
\bibitem [{\citenamefont {Gambini}\ \emph {et~al.}(2023)\citenamefont
  {Gambini}, \citenamefont {Olmedo},\ and\ \citenamefont
  {Pullin}}]{Gambini:2022hxr}%
  \BibitemOpen
  \bibfield  {author} {\bibinfo {author} {\bibfnamefont {Rodolfo}\ \bibnamefont
  {Gambini}}, \bibinfo {author} {\bibfnamefont {Javier}\ \bibnamefont
  {Olmedo}}, \ and\ \bibinfo {author} {\bibfnamefont {Jorge}\ \bibnamefont
  {Pullin}},\ }\enquote {\bibinfo {title} {{Quantum Geometry and Black
  Holes}},}\ \ (\bibinfo {year} {2023})\ \Eprint
  {http://arxiv.org/abs/2211.05621} {arXiv:2211.05621 [gr-qc]} \BibitemShut
  {NoStop}%
\bibitem [{\citenamefont {Lewandowski}\ \emph {et~al.}(2023)\citenamefont
  {Lewandowski}, \citenamefont {Ma}, \citenamefont {Yang},\ and\ \citenamefont
  {Zhang}}]{Lewandowski:2022zce}%
  \BibitemOpen
  \bibfield  {author} {\bibinfo {author} {\bibfnamefont {Jerzy}\ \bibnamefont
  {Lewandowski}}, \bibinfo {author} {\bibfnamefont {Yongge}\ \bibnamefont
  {Ma}}, \bibinfo {author} {\bibfnamefont {Jinsong}\ \bibnamefont {Yang}}, \
  and\ \bibinfo {author} {\bibfnamefont {Cong}\ \bibnamefont {Zhang}},\
  }\bibfield  {title} {\enquote {\bibinfo {title} {{Quantum Oppenheimer-Snyder
  and Swiss Cheese Models}},}\ }\href {\doibase 10.1103/PhysRevLett.130.101501}
  {\bibfield  {journal} {\bibinfo  {journal} {Phys. Rev. Lett.}\ }\textbf
  {\bibinfo {volume} {130}},\ \bibinfo {pages} {101501} (\bibinfo {year}
  {2023})},\ \Eprint {http://arxiv.org/abs/2210.02253} {arXiv:2210.02253
  [gr-qc]} \BibitemShut {NoStop}%
\bibitem [{\citenamefont {Alonso-Bardaji}\ and\ \citenamefont
  {Brizuela}(2024)}]{Alonso-Bardaji:2023vtl}%
  \BibitemOpen
  \bibfield  {author} {\bibinfo {author} {\bibfnamefont {Asier}\ \bibnamefont
  {Alonso-Bardaji}}\ and\ \bibinfo {author} {\bibfnamefont {David}\
  \bibnamefont {Brizuela}},\ }\bibfield  {title} {\enquote {\bibinfo {title}
  {{Spacetime geometry from canonical spherical gravity}},}\ }\href {\doibase
  10.1103/PhysRevD.109.044065} {\bibfield  {journal} {\bibinfo  {journal}
  {Phys. Rev. D}\ }\textbf {\bibinfo {volume} {109}},\ \bibinfo {pages}
  {044065} (\bibinfo {year} {2024})},\ \Eprint
  {http://arxiv.org/abs/2310.12951} {arXiv:2310.12951 [gr-qc]} \BibitemShut
  {NoStop}%
\bibitem [{\citenamefont {Giesel}\ \emph
  {et~al.}(2023{\natexlab{a}})\citenamefont {Giesel}, \citenamefont {Liu},
  \citenamefont {Singh},\ and\ \citenamefont {Weigl}}]{Giesel:2023hys}%
  \BibitemOpen
  \bibfield  {author} {\bibinfo {author} {\bibfnamefont {Kristina}\
  \bibnamefont {Giesel}}, \bibinfo {author} {\bibfnamefont {Hongguang}\
  \bibnamefont {Liu}}, \bibinfo {author} {\bibfnamefont {Parampreet}\
  \bibnamefont {Singh}}, \ and\ \bibinfo {author} {\bibfnamefont
  {Stefan~Andreas}\ \bibnamefont {Weigl}},\ }\bibfield  {title} {\enquote
  {\bibinfo {title} {{Generalized analysis of a dust collapse in effective loop
  quantum gravity: fate of shocks and covariance}},}\ }\href@noop {} {\
  (\bibinfo {year} {2023}{\natexlab{a}})},\ \Eprint
  {http://arxiv.org/abs/2308.10953} {arXiv:2308.10953 [gr-qc]} \BibitemShut
  {NoStop}%
\bibitem [{\citenamefont {Giesel}\ \emph
  {et~al.}(2023{\natexlab{b}})\citenamefont {Giesel}, \citenamefont {Liu},
  \citenamefont {Rullit}, \citenamefont {Singh},\ and\ \citenamefont
  {Weigl}}]{Giesel:2023tsj}%
  \BibitemOpen
  \bibfield  {author} {\bibinfo {author} {\bibfnamefont {Kristina}\
  \bibnamefont {Giesel}}, \bibinfo {author} {\bibfnamefont {Hongguang}\
  \bibnamefont {Liu}}, \bibinfo {author} {\bibfnamefont {Eric}\ \bibnamefont
  {Rullit}}, \bibinfo {author} {\bibfnamefont {Parampreet}\ \bibnamefont
  {Singh}}, \ and\ \bibinfo {author} {\bibfnamefont {Stefan~Andreas}\
  \bibnamefont {Weigl}},\ }\bibfield  {title} {\enquote {\bibinfo {title}
  {{Embedding generalized LTB models in polymerized spherically symmetric
  spacetimes}},}\ }\href@noop {} {\  (\bibinfo {year} {2023}{\natexlab{b}})},\
  \Eprint {http://arxiv.org/abs/2308.10949} {arXiv:2308.10949 [gr-qc]}
  \BibitemShut {NoStop}%
\bibitem [{\citenamefont {Alonso~Bardaj\'\i{}}(2023)}]{AlonsoBardaji:2023bww}%
  \BibitemOpen
  \bibfield  {author} {\bibinfo {author} {\bibfnamefont {Asier}\ \bibnamefont
  {Alonso~Bardaj\'\i{}}},\ }\emph {\bibinfo {title} {{Loop Quantum Gravity
  Effects on Spherical Black Holes. A Covariant Approach to Singularity
  Resolution}}},\ \href@noop {} {Ph.D. thesis},\ \bibinfo  {school} {U. Basque
  Country, Leioa} (\bibinfo {year} {2023})\BibitemShut {NoStop}%
\bibitem [{\citenamefont {Fazzini}\ \emph {et~al.}(2024)\citenamefont
  {Fazzini}, \citenamefont {Husain},\ and\ \citenamefont
  {Wilson-Ewing}}]{Fazzini:2023ova}%
  \BibitemOpen
  \bibfield  {author} {\bibinfo {author} {\bibfnamefont {Francesco}\
  \bibnamefont {Fazzini}}, \bibinfo {author} {\bibfnamefont {Viqar}\
  \bibnamefont {Husain}}, \ and\ \bibinfo {author} {\bibfnamefont {Edward}\
  \bibnamefont {Wilson-Ewing}},\ }\bibfield  {title} {\enquote {\bibinfo
  {title} {{Shell-crossings and shock formation during gravitational collapse
  in effective loop quantum gravity}},}\ }\href {\doibase
  10.1103/PhysRevD.109.084052} {\bibfield  {journal} {\bibinfo  {journal}
  {Phys. Rev. D}\ }\textbf {\bibinfo {volume} {109}},\ \bibinfo {pages}
  {084052} (\bibinfo {year} {2024})},\ \Eprint
  {http://arxiv.org/abs/2312.02032} {arXiv:2312.02032 [gr-qc]} \BibitemShut
  {NoStop}%
\bibitem [{\citenamefont {Zhang}\ \emph {et~al.}(2025)\citenamefont {Zhang},
  \citenamefont {Lewandowski}, \citenamefont {Ma},\ and\ \citenamefont
  {Yang}}]{Zhang:2024khj}%
  \BibitemOpen
  \bibfield  {author} {\bibinfo {author} {\bibfnamefont {Cong}\ \bibnamefont
  {Zhang}}, \bibinfo {author} {\bibfnamefont {Jerzy}\ \bibnamefont
  {Lewandowski}}, \bibinfo {author} {\bibfnamefont {Yongge}\ \bibnamefont
  {Ma}}, \ and\ \bibinfo {author} {\bibfnamefont {Jinsong}\ \bibnamefont
  {Yang}},\ }\bibfield  {title} {\enquote {\bibinfo {title} {{Black holes and
  covariance in effective quantum gravity}},}\ }\href {\doibase
  10.1103/PhysRevD.111.L081504} {\bibfield  {journal} {\bibinfo  {journal}
  {Phys. Rev. D}\ }\textbf {\bibinfo {volume} {111}},\ \bibinfo {pages}
  {L081504} (\bibinfo {year} {2025})},\ \Eprint
  {http://arxiv.org/abs/2407.10168} {arXiv:2407.10168 [gr-qc]} \BibitemShut
  {NoStop}%
\bibitem [{\citenamefont {Cafaro}\ and\ \citenamefont
  {Lewandowski}(2024)}]{Cafaro:2024vrw}%
  \BibitemOpen
  \bibfield  {author} {\bibinfo {author} {\bibfnamefont {Luca}\ \bibnamefont
  {Cafaro}}\ and\ \bibinfo {author} {\bibfnamefont {Jerzy}\ \bibnamefont
  {Lewandowski}},\ }\bibfield  {title} {\enquote {\bibinfo {title} {{Status of
  Birkhoff's theorem in polymerized semiclassical regime of Loop Quantum
  Gravity}},}\ }\href@noop {} {\  (\bibinfo {year} {2024})},\ \Eprint
  {http://arxiv.org/abs/2403.01910} {arXiv:2403.01910 [gr-qc]} \BibitemShut
  {NoStop}%
\bibitem [{\citenamefont {Lin}\ and\ \citenamefont
  {Zhang}(2024)}]{Lin:2024flv}%
  \BibitemOpen
  \bibfield  {author} {\bibinfo {author} {\bibfnamefont {Jianhui}\ \bibnamefont
  {Lin}}\ and\ \bibinfo {author} {\bibfnamefont {Xiangdong}\ \bibnamefont
  {Zhang}},\ }\bibfield  {title} {\enquote {\bibinfo {title} {{Effective
  four-dimensional loop quantum black hole with a cosmological constant}},}\
  }\href {\doibase 10.1103/PhysRevD.110.026002} {\bibfield  {journal} {\bibinfo
   {journal} {Phys. Rev. D}\ }\textbf {\bibinfo {volume} {110}},\ \bibinfo
  {pages} {026002} (\bibinfo {year} {2024})},\ \Eprint
  {http://arxiv.org/abs/2402.13638} {arXiv:2402.13638 [gr-qc]} \BibitemShut
  {NoStop}%
\bibitem [{\citenamefont {Cipriani}\ \emph {et~al.}(2024)\citenamefont
  {Cipriani}, \citenamefont {Fazzini},\ and\ \citenamefont
  {Wilson-Ewing}}]{Cipriani:2024nhx}%
  \BibitemOpen
  \bibfield  {author} {\bibinfo {author} {\bibfnamefont {Lorenzo}\ \bibnamefont
  {Cipriani}}, \bibinfo {author} {\bibfnamefont {Francesco}\ \bibnamefont
  {Fazzini}}, \ and\ \bibinfo {author} {\bibfnamefont {Edward}\ \bibnamefont
  {Wilson-Ewing}},\ }\bibfield  {title} {\enquote {\bibinfo {title}
  {{Gravitational collapse in effective loop quantum gravity: Beyond marginally
  bound configurations}},}\ }\href {\doibase 10.1103/PhysRevD.110.066004}
  {\bibfield  {journal} {\bibinfo  {journal} {Phys. Rev. D}\ }\textbf {\bibinfo
  {volume} {110}},\ \bibinfo {pages} {066004} (\bibinfo {year} {2024})},\
  \Eprint {http://arxiv.org/abs/2404.04192} {arXiv:2404.04192 [gr-qc]}
  \BibitemShut {NoStop}%
\bibitem [{\citenamefont {Zhang}\ \emph {et~al.}(2024)\citenamefont {Zhang},
  \citenamefont {Lewandowski}, \citenamefont {Ma},\ and\ \citenamefont
  {Yang}}]{Zhang:2024ney}%
  \BibitemOpen
  \bibfield  {author} {\bibinfo {author} {\bibfnamefont {Cong}\ \bibnamefont
  {Zhang}}, \bibinfo {author} {\bibfnamefont {Jerzy}\ \bibnamefont
  {Lewandowski}}, \bibinfo {author} {\bibfnamefont {Yongge}\ \bibnamefont
  {Ma}}, \ and\ \bibinfo {author} {\bibfnamefont {Jinsong}\ \bibnamefont
  {Yang}},\ }\bibfield  {title} {\enquote {\bibinfo {title} {{Black holes and
  covariance in effective quantum gravity: A solution without Cauchy
  horizons}},}\ }\href@noop {} {\  (\bibinfo {year} {2024})},\ \Eprint
  {http://arxiv.org/abs/2412.02487} {arXiv:2412.02487 [gr-qc]} \BibitemShut
  {NoStop}%
\bibitem [{\citenamefont {Lin}\ \emph {et~al.}(2024)\citenamefont {Lin},
  \citenamefont {Zhang},\ and\ \citenamefont {Bravo-Gaete}}]{Lin:2024beb}%
  \BibitemOpen
  \bibfield  {author} {\bibinfo {author} {\bibfnamefont {Jianhui}\ \bibnamefont
  {Lin}}, \bibinfo {author} {\bibfnamefont {Xiangdong}\ \bibnamefont {Zhang}},
  \ and\ \bibinfo {author} {\bibfnamefont {Mois\'es}\ \bibnamefont
  {Bravo-Gaete}},\ }\bibfield  {title} {\enquote {\bibinfo {title} {{Mass
  inflation and strong cosmic censorship conjecture in the covariant quantum
  black hole}},}\ }\href@noop {} {\  (\bibinfo {year} {2024})},\ \Eprint
  {http://arxiv.org/abs/2412.01448} {arXiv:2412.01448 [gr-qc]} \BibitemShut
  {NoStop}%
\bibitem [{\citenamefont {Yang}\ \emph {et~al.}(2025)\citenamefont {Yang},
  \citenamefont {Zhang},\ and\ \citenamefont {Ma}}]{Yang:2025ufs}%
  \BibitemOpen
  \bibfield  {author} {\bibinfo {author} {\bibfnamefont {Jinsong}\ \bibnamefont
  {Yang}}, \bibinfo {author} {\bibfnamefont {Cong}\ \bibnamefont {Zhang}}, \
  and\ \bibinfo {author} {\bibfnamefont {Yongge}\ \bibnamefont {Ma}},\
  }\bibfield  {title} {\enquote {\bibinfo {title} {{Covariant effective
  spacetimes of spherically symmetric electro-vacuum with a cosmological
  constant}},}\ }\href@noop {} {\  (\bibinfo {year} {2025})},\ \Eprint
  {http://arxiv.org/abs/2503.15157} {arXiv:2503.15157 [gr-qc]} \BibitemShut
  {NoStop}%
\bibitem [{\citenamefont {Fazzini}(2025{\natexlab{a}})}]{Fazzini:2025ysd}%
  \BibitemOpen
  \bibfield  {author} {\bibinfo {author} {\bibfnamefont {Francesco}\
  \bibnamefont {Fazzini}},\ }\bibfield  {title} {\enquote {\bibinfo {title}
  {{Gentle spaghettification in effective LQG dust collapse}},}\ }\href@noop {}
  {\  (\bibinfo {year} {2025}{\natexlab{a}})},\ \Eprint
  {http://arxiv.org/abs/2502.20187} {arXiv:2502.20187 [gr-qc]} \BibitemShut
  {NoStop}%
\bibitem [{\citenamefont {Liu}\ and\ \citenamefont {Qu}(2025)}]{Liu:2025fil}%
  \BibitemOpen
  \bibfield  {author} {\bibinfo {author} {\bibfnamefont {Hongguang}\
  \bibnamefont {Liu}}\ and\ \bibinfo {author} {\bibfnamefont {Dongxue}\
  \bibnamefont {Qu}},\ }\bibfield  {title} {\enquote {\bibinfo {title}
  {{Quantum induced shock dynamics in gravitational collapse: insights from
  effective models and numerical frameworks}},}\ }\href@noop {} {\  (\bibinfo
  {year} {2025})},\ \Eprint {http://arxiv.org/abs/2504.18462} {arXiv:2504.18462
  [gr-qc]} \BibitemShut {NoStop}%
\bibitem [{\citenamefont {Fazzini}\ and\ \citenamefont
  {Mehmood}(2025)}]{Fazzini:2025zrq}%
  \BibitemOpen
  \bibfield  {author} {\bibinfo {author} {\bibfnamefont {Francesco}\
  \bibnamefont {Fazzini}}\ and\ \bibinfo {author} {\bibfnamefont {Hassan}\
  \bibnamefont {Mehmood}},\ }\bibfield  {title} {\enquote {\bibinfo {title}
  {{On weak solutions in Einstein theory and beyond}},}\ }\href@noop {} {\
  (\bibinfo {year} {2025})},\ \Eprint {http://arxiv.org/abs/2505.01846}
  {arXiv:2505.01846 [gr-qc]} \BibitemShut {NoStop}%
\bibitem [{\citenamefont {Shi}\ \emph {et~al.}(2024)\citenamefont {Shi},
  \citenamefont {Zhang},\ and\ \citenamefont {Ma}}]{Shi:2024vki}%
  \BibitemOpen
  \bibfield  {author} {\bibinfo {author} {\bibfnamefont {Zijian}\ \bibnamefont
  {Shi}}, \bibinfo {author} {\bibfnamefont {Xiangdong}\ \bibnamefont {Zhang}},
  \ and\ \bibinfo {author} {\bibfnamefont {Yongge}\ \bibnamefont {Ma}},\
  }\bibfield  {title} {\enquote {\bibinfo {title} {{Higher-dimensional quantum
  Oppenheimer-Snyder model}},}\ }\href {\doibase 10.1103/PhysRevD.110.104074}
  {\bibfield  {journal} {\bibinfo  {journal} {Phys. Rev. D}\ }\textbf {\bibinfo
  {volume} {110}},\ \bibinfo {pages} {104074} (\bibinfo {year} {2024})},\
  \Eprint {http://arxiv.org/abs/2408.15821} {arXiv:2408.15821 [gr-qc]}
  \BibitemShut {NoStop}%
\bibitem [{\citenamefont {Israel}(1966)}]{Israel:1966rt}%
  \BibitemOpen
  \bibfield  {author} {\bibinfo {author} {\bibfnamefont {W.}~\bibnamefont
  {Israel}},\ }\bibfield  {title} {\enquote {\bibinfo {title} {{Singular
  hypersurfaces and thin shells in general relativity}},}\ }\href {\doibase
  10.1007/BF02710419} {\bibfield  {journal} {\bibinfo  {journal} {Nuovo Cim.
  B}\ }\textbf {\bibinfo {volume} {44S10}},\ \bibinfo {pages} {1} (\bibinfo
  {year} {1966})},\ \bibinfo {note} {[Erratum: Nuovo Cim.B 48, 463
  (1967)]}\BibitemShut {NoStop}%
\bibitem [{\citenamefont {Friedman}\ \emph {et~al.}(1997)\citenamefont
  {Friedman}, \citenamefont {Louko},\ and\ \citenamefont
  {Winters-Hilt}}]{Friedman:1997fu}%
  \BibitemOpen
  \bibfield  {author} {\bibinfo {author} {\bibfnamefont {John~L.}\ \bibnamefont
  {Friedman}}, \bibinfo {author} {\bibfnamefont {Jorma}\ \bibnamefont {Louko}},
  \ and\ \bibinfo {author} {\bibfnamefont {Stephen~N.}\ \bibnamefont
  {Winters-Hilt}},\ }\bibfield  {title} {\enquote {\bibinfo {title} {{Reduced
  phase space formalism for spherically symmetric geometry with a massive dust
  shell}},}\ }\href {\doibase 10.1103/PhysRevD.56.7674} {\bibfield  {journal}
  {\bibinfo  {journal} {Phys. Rev. D}\ }\textbf {\bibinfo {volume} {56}},\
  \bibinfo {pages} {7674--7691} (\bibinfo {year} {1997})},\ \Eprint
  {http://arxiv.org/abs/gr-qc/9706051} {arXiv:gr-qc/9706051} \BibitemShut
  {NoStop}%
\bibitem [{\citenamefont {Casado-Turri\'on}(2023)}]{Casado-Turrion:2023jba}%
  \BibitemOpen
  \bibfield  {author} {\bibinfo {author} {\bibfnamefont {Adri\'an}\
  \bibnamefont {Casado-Turri\'on}},\ }\emph {\bibinfo {title} {{Compact objects
  in modified gravity: junction conditions and other viability criteria}}},\
  \href@noop {} {Ph.D. thesis},\ \bibinfo  {school} {Madrid U.} (\bibinfo
  {year} {2023}),\ \Eprint {http://arxiv.org/abs/2312.03757} {arXiv:2312.03757
  [gr-qc]} \BibitemShut {NoStop}%
\bibitem [{\citenamefont {Han}\ and\ \citenamefont {Liu}(2024)}]{Han:2022rsx}%
  \BibitemOpen
  \bibfield  {author} {\bibinfo {author} {\bibfnamefont {Muxin}\ \bibnamefont
  {Han}}\ and\ \bibinfo {author} {\bibfnamefont {Hongguang}\ \bibnamefont
  {Liu}},\ }\bibfield  {title} {\enquote {\bibinfo {title} {{Covariant
  $\bar\mu$-scheme effective dynamics, mimetic gravity, and nonsingular black
  holes: Applications to spherically symmetric quantum gravity}},}\ }\href
  {\doibase 10.1103/PhysRevD.109.084033} {\bibfield  {journal} {\bibinfo
  {journal} {Phys. Rev. D}\ }\textbf {\bibinfo {volume} {109}},\ \bibinfo
  {pages} {084033} (\bibinfo {year} {2024})},\ \Eprint
  {http://arxiv.org/abs/2212.04605} {arXiv:2212.04605 [gr-qc]} \BibitemShut
  {NoStop}%
\bibitem [{\citenamefont {Belfaqih}\ \emph {et~al.}(2024)\citenamefont
  {Belfaqih}, \citenamefont {Bojowald}, \citenamefont {Brahma},\ and\
  \citenamefont {Duque}}]{Belfaqih:2024vfk}%
  \BibitemOpen
  \bibfield  {author} {\bibinfo {author} {\bibfnamefont {Idrus~Husin}\
  \bibnamefont {Belfaqih}}, \bibinfo {author} {\bibfnamefont {Martin}\
  \bibnamefont {Bojowald}}, \bibinfo {author} {\bibfnamefont {Suddhasattwa}\
  \bibnamefont {Brahma}}, \ and\ \bibinfo {author} {\bibfnamefont {Erick~I.}\
  \bibnamefont {Duque}},\ }\bibfield  {title} {\enquote {\bibinfo {title}
  {{Black holes in effective loop quantum gravity: Covariant holonomy
  modifications}},}\ }\href@noop {} {\  (\bibinfo {year} {2024})},\ \Eprint
  {http://arxiv.org/abs/2407.12087} {arXiv:2407.12087 [gr-qc]} \BibitemShut
  {NoStop}%
\bibitem [{\citenamefont {Fiamberti}\ and\ \citenamefont
  {Menotti}(2008)}]{Fiamberti:2007qb}%
  \BibitemOpen
  \bibfield  {author} {\bibinfo {author} {\bibfnamefont {Francesco}\
  \bibnamefont {Fiamberti}}\ and\ \bibinfo {author} {\bibfnamefont {Pietro}\
  \bibnamefont {Menotti}},\ }\bibfield  {title} {\enquote {\bibinfo {title}
  {{Reduced Hamiltonian for intersecting shells}},}\ }\href {\doibase
  10.1016/j.nuclphysb.2007.11.003} {\bibfield  {journal} {\bibinfo  {journal}
  {Nucl. Phys. B}\ }\textbf {\bibinfo {volume} {794}},\ \bibinfo {pages}
  {512--537} (\bibinfo {year} {2008})},\ \Eprint
  {http://arxiv.org/abs/0708.2868} {arXiv:0708.2868 [hep-th]} \BibitemShut
  {NoStop}%
\bibitem [{\citenamefont {Fazzini}(2025{\natexlab{b}})}]{Fazzini:2025hsf}%
  \BibitemOpen
  \bibfield  {author} {\bibinfo {author} {\bibfnamefont {Francesco}\
  \bibnamefont {Fazzini}},\ }\bibfield  {title} {\enquote {\bibinfo {title}
  {{Non-uniqueness of the shockwave dynamics in effective loop quantum
  gravity}},}\ }\href@noop {} {\  (\bibinfo {year} {2025}{\natexlab{b}})},\
  \Eprint {http://arxiv.org/abs/2502.03003} {arXiv:2502.03003 [gr-qc]}
  \BibitemShut {NoStop}%
\bibitem [{\citenamefont {Hajicek}\ and\ \citenamefont
  {Kijowski}(1998)}]{Hajicek:1997va}%
  \BibitemOpen
  \bibfield  {author} {\bibinfo {author} {\bibfnamefont {P.}~\bibnamefont
  {Hajicek}}\ and\ \bibinfo {author} {\bibfnamefont {J.}~\bibnamefont
  {Kijowski}},\ }\bibfield  {title} {\enquote {\bibinfo {title} {{Lagrangian
  and Hamiltonian formalism for discontinuous fluid and gravitational
  field}},}\ }\href {\doibase 10.1103/PhysRevD.57.914} {\bibfield  {journal}
  {\bibinfo  {journal} {Phys. Rev. D}\ }\textbf {\bibinfo {volume} {57}},\
  \bibinfo {pages} {914--935} (\bibinfo {year} {1998})},\ \bibinfo {note}
  {[Erratum: Phys.Rev.D 61, 129901 (2000)]},\ \Eprint
  {http://arxiv.org/abs/gr-qc/9707020} {arXiv:gr-qc/9707020} \BibitemShut
  {NoStop}%
\bibitem [{\citenamefont {Zhang}\ and\ \citenamefont
  {Cao}(2025)}]{Zhang:2025ccx}%
  \BibitemOpen
  \bibfield  {author} {\bibinfo {author} {\bibfnamefont {Cong}\ \bibnamefont
  {Zhang}}\ and\ \bibinfo {author} {\bibfnamefont {Zhoujian}\ \bibnamefont
  {Cao}},\ }\bibfield  {title} {\enquote {\bibinfo {title} {{Covariant dynamics
  from static spherically symmetric geometries}},}\ }\href@noop {} {\
  (\bibinfo {year} {2025})},\ \Eprint {http://arxiv.org/abs/2506.09540}
  {arXiv:2506.09540 [gr-qc]} \BibitemShut {NoStop}%
\end{thebibliography}
%

\end{document}